\documentclass[bibyear]{aa}
\usepackage{graphics,graphicx,graphbox}
\usepackage{txfonts}
\usepackage{amsmath}	
\usepackage{subcaption}
\usepackage{tabularx}
\usepackage{footnote}
\usepackage{pdflscape}
\usepackage{url}
\usepackage{booktabs,caption}
\usepackage[export]{adjustbox} 
\usepackage{natbib}
\usepackage{ulem}
\usepackage[flushleft]{threeparttable}
\usepackage{multirow}
\usepackage{gensymb}
\usepackage{scrextend}   
\usepackage[colorlinks=true, linkcolor=blue, citecolor=blue, draft=false]{hyperref}
\usepackage{xcolor}

\begin{document}

   \title{Peculiar velocities of galaxy clusters in a IllustrisTNG simulation based on mock observations of Sunyaev-Zel'dovich effects}
   \titlerunning{Peculiar velocities  based on the kSZ effect of TNG300-3 galaxy clusters}
   
   \author{Kaiyi Du\inst{1,2}
          \and
          Yong Shi\inst{1,2}\fnmsep\thanks{yong@nju.edu.cn}
          \and
          Tao Wang\inst{1,2}
          \and
          Chenggang Shu\inst{3}
          }

   \institute{School of Astronomy and Space Science, Nanjing University, Nanjing 210093, People’s Republic of China
         \and
             Key Laboratory of Modern Astronomy and Astrophysics (Nanjing University), Ministry of Education, Nanjing 210093, People’s Republic of China
        \and 
            Shanghai Key Lab for Astrophysics, Shanghai Normal University, Shanghai 200234, China\\
             }

   \date{Received July 30, 2024; accepted January 24, 2025}
   
   \abstract{
   Galaxy clusters are the largest self-gravitational systems in the Universe. They are valuable probes of the structure growth of the Universe when we estimate their peculiar velocities with the kinematic Sunyaev-Zel'dovich (kSZ) effects. We investigate whether there is a systematic offset between the peculiar velocities ($v_{\rm z,kSZ}$) estimated with the kSZ effect and the true velocities of halos. We first created mocks of the 2D maps of SZ effects in seven frequency bands for galaxy clusters spanning a broad range of masses in TNG 300-3. We then derived the line-of-sight (LOS) peculiar velocities of galaxy clusters by applying an analytical formula to fit the spectra of the SZ effect. We find that the analytical formula-fitting method tends to overestimate the peculiar velocities of galaxy clusters, regardless of whether they approach us or recede from us. However, when galaxy clusters larger than 500 km/s were excluded, the slopes of the relations between $v_{\rm z,kSZ}$ and the real LOS velocities of galaxy clusters ($v_{\rm z,halo}$) were consistent with unity within the errors. Additionally, we further accounted for observational noise for different-aperture telescopes under different precipitable water vapor. The slopes for most cases are consistent with unity within the errors, and the errors of the slopes slightly increase with higher observation noise. The $|v_{\rm z,kSZ}-v_{\rm z,halo}|$ exhibits no obvious trend with increasing concentration and $M_{\rm 200}$. It is noteworthy that the $|v_{\rm z,kSZ}-v_{\rm z,halo}|$ is relatively high for galaxy clusters with strong active galactic nucleus feedback and high star formation rates. This indicates that these physical processes affect the cluster dynamic state and may reflect on the peculiar velocity estimated from the kSZ effect.
   }
   
   \maketitle



\section{Introduction}
\label{sec:intro}

Galaxy clusters are the largest self-gravitationally bound systems in the Universe. They contain large quantities of gas, member galaxies, and dark matter (DM). In the hierarchical model of cosmic structure formation, galaxy clusters form and evolve through the merging of galaxy groups and clusters and continuous accretion of surrounding gas \citep{Kravtsov2012}. Galaxy clusters offer powerful probes of the large-scale structure and velocity field of the Universe \citep{Carlstrom2002,Allen2011,PlanckCollaboration2016} and provide laboratories for understanding the interaction of galaxies, active galactic nuclei (AGN), and star formation in different environments.

The traditional method for studying the intracluster medium (ICM) within galaxy clusters is through X-ray emissions from the hot gas. However, an X-ray detection becomes increasingly challenging at $z \gtrsim 1$ because the X-ray flux decreases rapidly with increasing $z$ \citep{vanMarrewijk2024}. Additionally, many existing X-ray observations are hampered by a low spectral resolution, which makes it difficult for X-ray observations to provide information about the motion of hot gas in clusters. In contrast, the Sunyaev-Zel'dovich (SZ) effect provides a unique and redshift-independent method of probing the ICM. The SZ effect contains the thermal-SZ (tSZ) effect \citep{Sunyaev1972}, characterized by the distortion of the cosmic microwave background (CMB) spectrum through inverse-Compton scattering of CMB photons by electrons in the ICM, and the kinematic-SZ (kSZ) effect \citep{Sunyaev1980}, arising from the scattering of CMB photons by the electrons in ICM due to the bulk motion of the cluster. The tSZ effect is proportional to the line-of-sight (LOS) integral of the electron pressure: $\frac{\Delta T_{tSZ}}{T_{CMB}}=f(x)y=f(x)\int n_e\frac{k_B T_e}{m_e c^2} \sigma_Tdl$ \footnote{$f(x)=(x\frac{e^{x}+1}{e^{x}-1}-4)(1+\delta_{\rm SZE}(x,T_{e}))$, where $\delta_{\rm SZE}(x,T_{e})$ is the relativistic correction to the frequency dependence. $x = \frac{h\nu}{k_{\rm B}T_{\rm CMB}}$ is the dimensionless frequency, with the observed frequency $\nu$, Planck constant $h$, Boltzmann’s constant $k_{\rm B}$, and the CMB temperature $T_{\rm CMB}$.},
while the kSZ effect is related to the integrated LOS electron density and the gas velocity with respect to the CMB reference frame: $\frac{\Delta T_{kSZ}}{T_{CMB}}=-\tau_e \frac{v_{pec}}{c}$ \citep{Carlstrom2002,PlanckCollaboration2016}. 

Compared to the tSZ effect, the kSZ effect is much weaker and difficult to detect unless the gas velocity reaches a few thousandths of the speed of light \citep{Birkinshaw1999}. Therefore, the direct observation of the kSZ signal in individual clusters requires high-sensitivity observations. The kSZ signal was only detected for a few individual galaxy clusters \citep{Sayers2013,Adam2017} or through statistical measurements \citep{Hand2012,PlanckCollaboration2016a,Tanimura2021,Chen2022}. This limits our understanding of the cosmic velocity field. Upcoming telescopes such as the Cerro-Chajnantor Atacama Telescope prime (CCAT-prime; \citealt{Parshley2018,Chapman2022,CPC2023}) and the Leighton Chajnantor Telescope (LCT; \citealt{Yao2023}) are expected to obtain measurements of the SZ effect in more than three bands and to detect the faint kSZ effect to trace the peculiar velocities of a large number of galaxy clusters \citep{Terry2019}. These measurements can provide us maps of the cosmic velocity field and powerful constraints on the structure growth of the Universe \citep{Okumura2022,Alonso2016,Adams2017,Adams2020,Turner2023, Lai2023,Shi2024}. If clusters were ideally spherically symmetric systems with minimal internal motions, then the kSZ effect would provide a relatively straightforward method of measuring their peculiar velocities. Nonetheless, galaxy clusters are complex systems with significant internal flows \citep{Ricker2001,Sun2002} driven by mergers and physical processes such as star formation and AGN feedback. Consequently, it becomes imperative to study the bias of the peculiar velocities that are estimated through the kSZ effect. The deviation of the slope of the $v_{\rm z,kSZ}$-$v_{\rm z,halo}$ relation from unity determines how precisely the kSZ effect can estimate the true peculiar velocity and constrain the cosmological parameters.

Only a few previous studies have focused on how accurate the peculiar velocity from the kSZ effect is. \citet{Nagai2003} used a high-resolution simulation of a galaxy cluster with a complex internal structure and motion to investigate the errors in the peculiar velocity estimated through the kSZ effect. They found a dispersion in the peculiar velocity estimated through the kSZ effect of $50-100$~$ \rm km\ s^{-1}$, and the velocity averaged over the kSZ map over the cluster virial radius provided an unbiased estimate. However, \citet{Nagai2003} did not simulate the observations of the SZ effect at different frequencies, and they only studied one single galaxy cluster. \citet{Wang2021} estimated the peculiar velocities using deep neural networks based on Magneticum simulations and found an average uncertainty of the peculiar velocity estimated through the kSZ of about 25\%. Although the sample of galaxy clusters in this work is large, the results were based on the temperature distortion maps of the SZ effect instead of mock observations that mimic real observations.

We carried out more realistic mocks of the SZ maps at seven frequencies for a sample of 155 galaxy clusters with masses higher than $10^{13.5}$~$\rm M_{\odot}$ and redshift ranges from $0.2-1.0$. Based on the mock observations, we obtained the LOS peculiar velocity ($v_{\rm z,kSZ}$) via the kSZ effect for each cluster and compared it with the real peculiar velocity ($v_{\rm z,halo}$) to study the bias of $v_{\rm z,kSZ}$.
We present the sample of galaxy clusters from the TNG300-3 simulation run of IllustrisTNG \citep{Nelson2019} in Section~\ref{sec:sample} and the procedures of simulated observations of the SZ effect in Section~\ref{sec:mock}. Section~\ref{sec:results} shows the main results of this work and compares the peculiar velocities estimated via the kSZ effect with the real peculiar velocities of galaxy clusters. We also present the cases with observation noise. In Section~\ref{sec:discussion}, we discuss the impacts of the galaxy cluster properties on the ${v}_{\rm z,kSZ}$. Section~\ref{sec:conclusion} concludes our work. Throughout this work, we assume $H_0=70{\rm km\ s^{-1}\ Mpc^{-1}}$, $\Omega_0=0.3$, and $\Omega_{\Lambda}=0.7$.

\begin{table*}
    \centering
    \caption{ PWV and the observation noise (rms) of a one-hour on-source observation for each pointing with 10-meter, 30-meter, and 100-meter radio telescopes at seven frequency bands, with an elevation of 45 degrees and under different PWV at CBI.}
	\label{tab:1}
    \begin{tabular}{ccccccccc}
    \hline
     Telescope Diameter (m) & PWV (mm) & & & & rms ($\mu$Jy/beam) & & & \\
     & & 90 GHz & 150 GHz & 230 GHz & 260 GHz & 345 GHz & 460 GHz & 650 GHz \\
     \hline
     10 & 0.5 & 127.7 & 127.7 & 175.7 & 192.1 & 368.2 & 1676.0 & 2871.0 \\
        & 1.0 & 134.0 & 156.3 & 239.2 & 277.3 & 626.3 & 4127.6 & 8347.7 \\
        & 1.5 & 143.5 & 182.2 & 305.8 & 364.5 & 913.6 & 8258.1 & 20103.6 \\
        & 2.0 & 153.1 & 212.1 & 379.5 & 457.3 & 1238.6 & 15194.5 & 45085.7 \\
     \hline
     30 & 0.5 & 14.2 & 14.2 & 19.5 & 21.3 & 40.9 & 186.2 & 319.0 \\
        & 1.0 & 14.9 & 17.4 & 26.6 & 30.8 & 69.6 & 458.5 & 927.5 \\
        & 1.5 & 15.9 & 20.2 & 34.0 & 40.5 & 101.5 & 917.6 & 2233.7 \\
        & 2.0 & 17.0 & 23.6 & 42.2 & 50.8 & 137.6 & 1688.3 & 5009.5 \\
    \hline
    100 & 0.5 & 1.3 & 1.3 & 1.8 & 1.9 & 3.7 & 16.8 & 28.7 \\
        & 1.0 & 1.3 & 1.6 & 2.4 & 2.8 & 6.3 & 41.3 & 83.5 \\
        & 1.5 & 1.4 & 1.8 & 3.1 & 3.6 & 9.1 & 82.6 & 201.0 \\
        & 2.0 & 1.5 & 2.1 & 3.8 & 4.6 & 12.4 & 151.9 & 450.9 \\
    \hline
    \end{tabular}
\end{table*}

\begin{figure}
    \includegraphics[width=1.1\columnwidth]{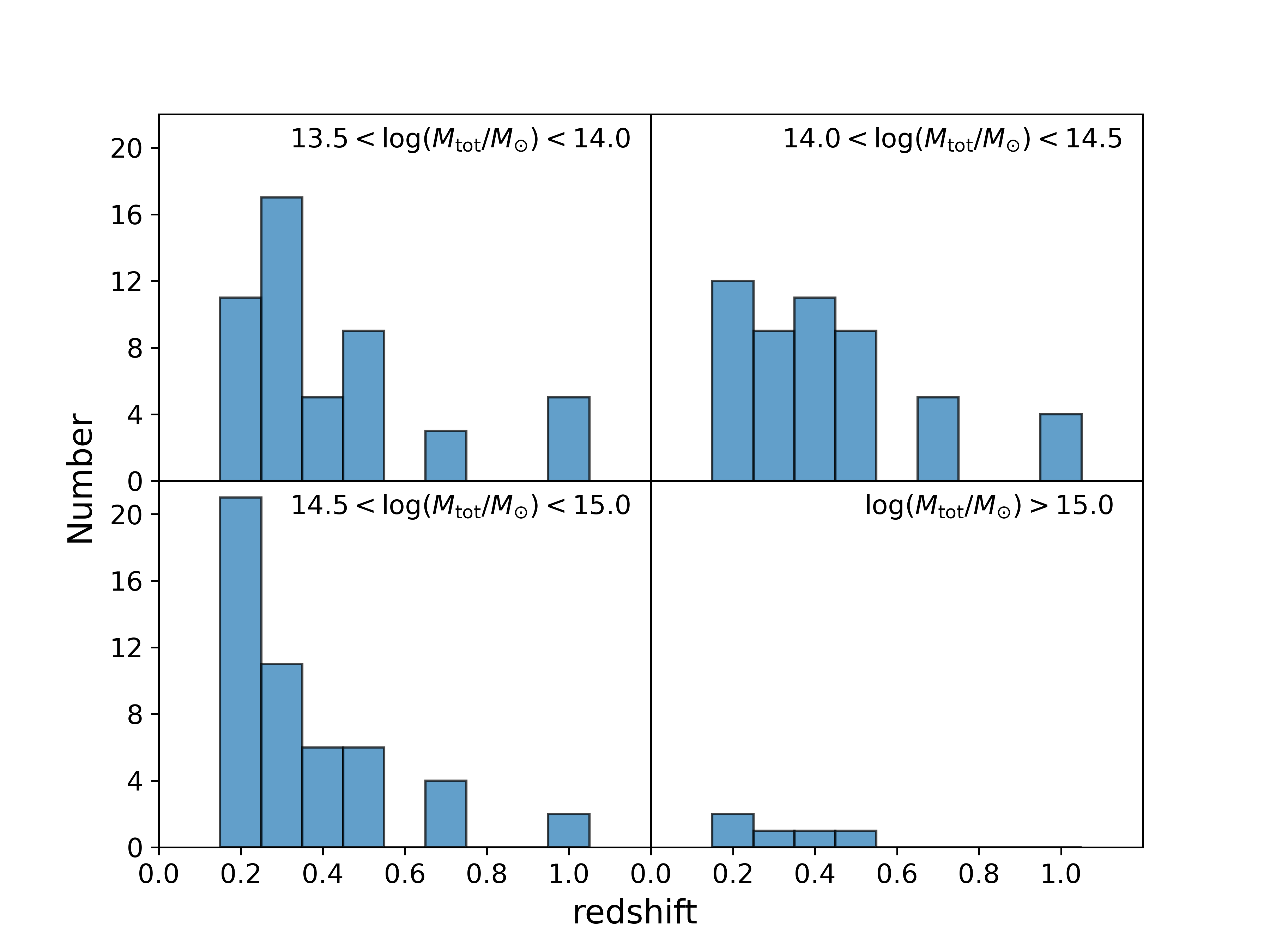}
    \caption{ Redshift distributions of the chosen galaxy clusters for the mass ranges$10^{13.5} \sim 10^{14.0}$~$\rm M_{\odot}$, $10^{14.0} \sim 10^{14.5}$~$\rm M_{\odot}$, $10^{14.5} \sim 10^{15.0}$~$\rm M_{\odot}$, and $>10^{15.0}$~$\rm M_{\odot}$. The whole sample includes 155 galaxy clusters, the first three mass ranges contain 50 sources each, and the most massive range only contains 5 sources.}
    \label{fig:sample}
\end{figure}

\section{Method}
\label{sec:method}

\begin{figure*}
    \centering
    \begin{subfigure}[b]{\linewidth}
        \centering     \includegraphics[width=\textwidth,keepaspectratio]{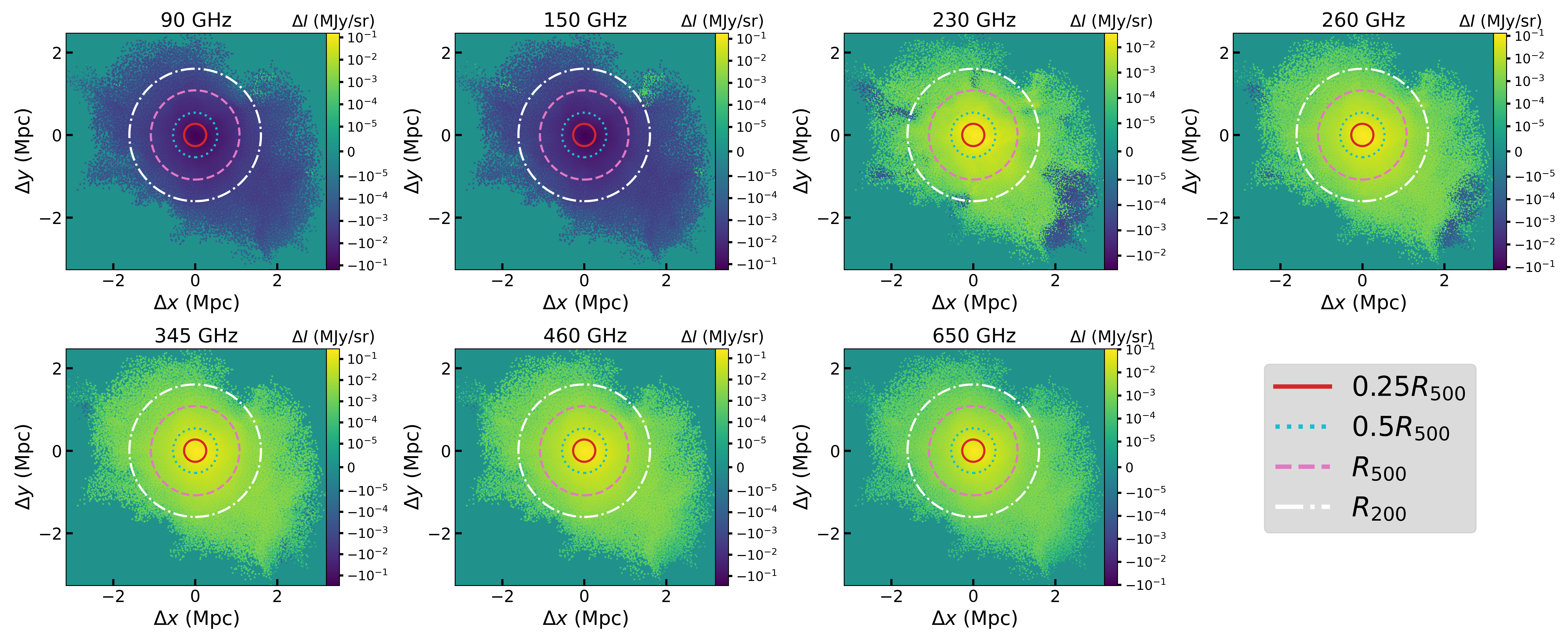}
        \caption{}
        \label{fig:SZmap}
    \end{subfigure}
    
    \begin{subfigure}[b]{\linewidth}
        \centering
        \includegraphics[width=\textwidth]{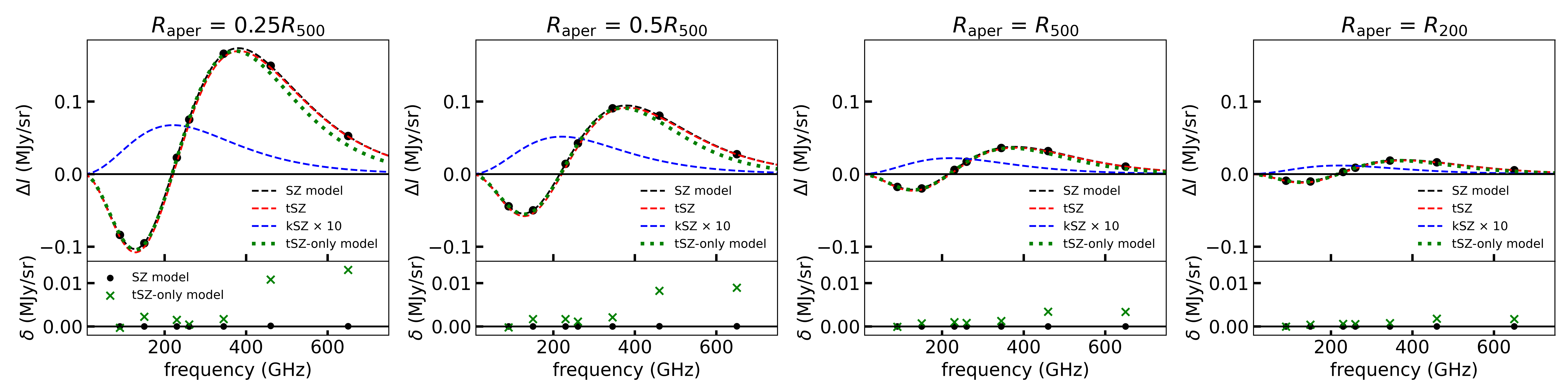}
        \caption{}
        \label{fig:SZspectrum}
    \end{subfigure}
    \caption{SZ effect of a galaxy cluster at various frequencies and the corresponding SZ spectra. Panel (a): Examples of the SZ effect of a galaxy cluster at various frequencies: 90~GHz, 150~GHz, 230~GHz, 260~GHz, 345~GHz, 460~GHz, and 650~GHz. The circles represent four different aperture sizes for photometry. The solid red circle corresponds to $0.25R_{\rm 500}$, the dotted cyan circle to $0.5R_{\rm 500}$, the dashed pink circle to $R_{\rm 500}$, and the dash-dotted white circle to $R_{\rm 200}$. Panel (b): SZ spectra of the galaxy cluster in panel~(a) measured at the cluster center using different aperture sizes of $0.25R_{\rm 500}$, $0.5R_{\rm 500}$, $R_{\rm 500}$, and $R_{\rm 200}$, corresponding to the circles in the SZ maps in panel~(a). The black dots represent the measured fluxes of the SZ effect at different frequencies. We fit the black dots with an analytic formula for the SZ effect \citep{Nozawa2006}. The dashed black lines show the results of fitting the total SZ signal, and the dashed red and blue lines represent the contributions from the tSZ and kSZ effects, respectively. The dotted green line corresponds to the best-fit result of the tSZ-only model. The residuals of the best-fit SZ model and the tSZ-only model with respect to the data are shown as black dots and green crosses in the bottom panels.}
    \label{fig:SZexample}
\end{figure*}

\begin{figure}
	\includegraphics[width=1.0\columnwidth]{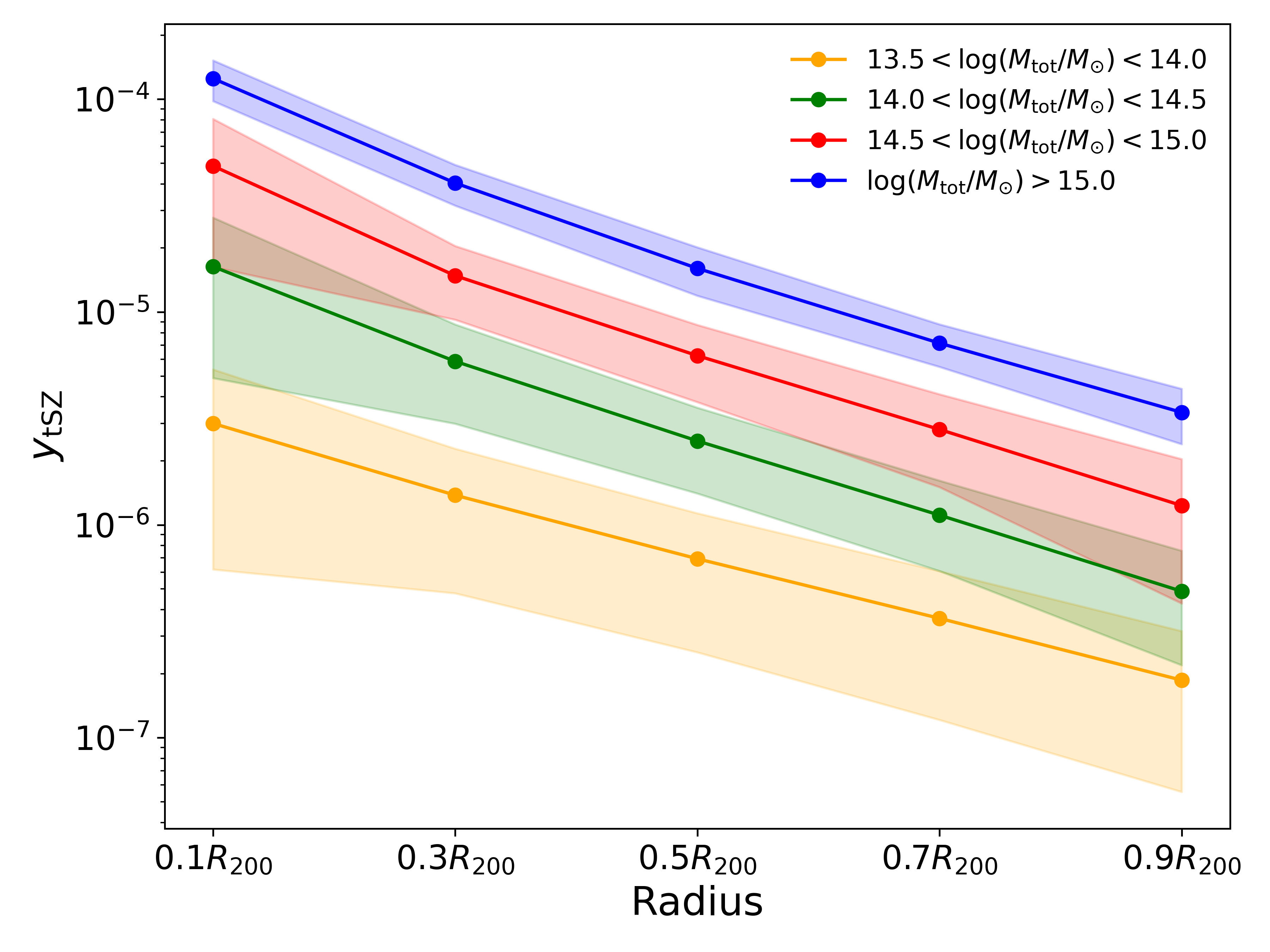}
    \caption{Radial profiles of the Compton $y_{\rm tSZ}$ parameter for the four mass bins of our sample. The radii are normalized to $R_{200}$ of the clusters. The lines represent the mean radial profile of $y_{\rm tSZ}$, and the shaded regions indicate the scatter.}
    \label{fig:ytSZ}
\end{figure}

\subsection{Sample}
\label{sec:sample}

We used data from IllustrisTNG \citep{Nelson2019}, a cosmological magnetohydrodynamical simulation that models the evolution of dark matter, gas, stars, and supermassive black holes
\citep{Springel2010,Pakmor2016}. We focused on the SZ effect of the whole galaxy cluster, whose typical size is a few megaparsec. We used the TNG300-3 simulation run with a large volume of $302.6^{3}$ Mpc but a low baryonic and DM resolution of $1.1\times10^{7}$~$\rm M_{\odot}$ and $5.9\times10^{7}$~$\rm M_{\odot}$. We selected a large sample of subhalos with masses higher than $10^{13.5}$~$\rm M_{\odot}$ and \texttt{primary$\_$flag} = 1, indicating that the halo is either the most massive or primary subhalo of the friends-of-friends (FoF) halo. The redshift range was set to be between 0.2 and 1. The sample contained 155 galaxy clusters in total, with 50 objects in each mass range of $10^{13.5} \sim 10^{14.0}$~$\rm M_{\odot}$, $10^{14.0} \sim 10^{14.5}$~$\rm M_{\odot}$, and $10^{14.5} \sim 10^{15.0}$~$\rm M_{\odot}$, along with five objects with masses higher than $10^{15.0}$~$\rm M_{\odot}$. We only chose subhalos from snapshots of the TNG300-3 with full information ($z = 0.2, 0.3, 0.4, 0.5, 0.7, 1.0$). Our clusters followed the redshift distribution of the subhalos in each mass range, which is shown in Figure~\ref{fig:sample}.

\subsection{Mock observations of the SZ effect}
\label{sec:mock}
To investigate whether there is a systematic error in the peculiar velocity estimated from the kSZ effect, we needed to simulate observations of the SZ effect of galaxy clusters at different frequencies. We used \texttt{SZpack} \citep{Chluba2012,Chluba2013}, a numerical library allowing fast and precise computation of the SZ signals for each gas cell of the galaxy clusters in TNG300-3 and projected it onto the x-y plane.

To obtain the SZ signals at different frequencies, we needed several fundamental physical quantities of each gas particle as the input to \texttt{SZpack}, including the optical depth ($\tau$), the gas temperatures ($kT$), and the velocities of gas particles ($v_{\rm gas}$). The gas velocities are directly given in the TNG300-3 dataset. The optical depths were derived from
\begin{equation}
    \tau = \sigma_{\rm T}\int{n_{\rm e}}dl,
\end{equation} 
where $\sigma_{T}$ is the Thomson cross-section
and $n_{\rm e}$ is the electron number densities. $n_{\rm e}$ is derived from the electron abundance $x_{\rm e}$, the hydrogen mass fraction $X_{\rm H}=0.76$, and the density of gas particles following the equation 
\begin{equation}
    n_{\rm e} = x_{\rm e} \times n_{\rm H},
\end{equation}
where $n_{\rm H} = X_{\rm H} *\rho/m_{\rm p}$ and $x_{\rm e}=n_{\rm e}/n_{\rm H}$ is given by \texttt{PartType0} snapshot fields of the TNG dataset.
The gas temperatures were derived from $x_{\rm e}$ and the internal energy $u$ following 
\begin{equation}
    T = (\gamma-1)\times \frac{u}{k_{\rm B}}\times \frac{UnitEnergy}{UnitMass}*\mu,
\end{equation}
where the adiabatic index $\gamma$ is $5/3$ and the Boltzmann constant $k_{\rm B}$ is in centimeter–gram–second units. $UnitMass$ and $UnitEnergy$ are code units, and $\frac{UnitEnergy}{UnitMass} = 10^{10}$. $u$ can also be found in \texttt{PartType0} snapshot fields of the TNG dataset. 
$\mu$ is the mean molecular weight, 
\begin{equation}
    \mu = \frac{4}{1+3 X_{\rm H}+4 X_{\rm H} x_{\rm e}}\times m_{\rm p}.
\end{equation}
We converted all quantities from comoving units into physical units.

We simulated the SZ signal at seven frequently used bands, including 90~GHz (frequency band of MUSTANG-2 camera on the Green Bank Telescope, GBT; \citealt{Dicker2014}), 150~GHz, 260~GHz (frequency bands of the NIKA2 (New IRAM KID Arrays 2) instrument on the IRAM 30M; \citealt{Adam2018,Catalano2018}),  and 230~GHz, 345~GHz, 460~GHz, and 650~GHz (frequency bands of the LCT; \citealt{Yao2023}).

We also studied the peculiar velocities obtained from the kSZ effect under certain observational noises. 
We estimated the observation noise for telescopes with assumed conditions, and we list them in Table~\ref{tab:1}. We considered one-hour on-source observations for each pointing of every galaxy cluster with 10-meter, 30-meter, and 100-meter radio telescopes at Chajnantor Plateau (CBI) under different precipitable water vapor (PWV). The observation noise was obtained following the equations in \citet{He2022},
\begin{equation}
    T_{\rm sky}=T_{\rm atm}(1-e^{-\tau_{\rm sky}}),
\end{equation}
where $T_{\rm sky}$ is the sky temperature, $T_{\rm atm}$ is the representative atmosphere temperature, and $\tau_{\rm sky}$ is the sky opacity. We adopted $T_{\rm atm}=273\ \rm K$, an average atmospheric temperature for the CBI \citep{Cortes2020}.  In this equation, $e^{-\tau_{\rm sky}}=\theta_{\rm trans}$  is the transmission, for which we adopted the transmission of the atmosphere at the CBI\footnote{\url{https://www.apex-telescope.org/sites/chajnantor/atmosphere/}}. Furthermore, $T_{sys}$ can be calculated following the equation in \citet{He2022},
\begin{equation}
    T_{\rm sys} \approx \frac{1}{e^{-\tau_{\rm sky}}}(T_{\rm rx}+T_{\rm sky}),
\end{equation}
in which $T_{\rm rx}$ is the receiver temperature. It is typically 10 to 20 K, so we assumed $T_{\rm rx}=10\ \rm K$ here.
The system equivalent flux density (SEFD) in $\rm Jy$ is
\begin{equation}
    {\rm SEFD} = \frac{T_{\rm sys}}{G},
\end{equation}
where $G$ is the antenna gain, $G=A_{\rm eff}/2k_{\rm B}$. Here, $A_{\rm eff}$ is the effective area of the antenna, and $A_{\rm eff} = \eta_{\rm A}\frac{\pi D^2}{4}$, where $\eta_{\rm A}\sim 0.7$ for a well-designed telescope, and $D$ is the diameter of the telescope.
Then, the rms noise in $\rm Jy/beam$ of the observation for a single-dish telescope is
\begin{equation}
    rms \approx \frac{c\times {\rm SEFD}}{\sqrt{\Delta \nu t_{\rm int}}} \approx \frac{c T_{\rm sys}}{G \sqrt{\Delta \nu t_{\rm int}}},
\end{equation}
where the $c$ accounts for the quantization and correlator efficiencies, and we assumed $c=0.9$. $\Delta \nu$ is the frequency bandwidth and $t_{\rm int}$ is the integration time of the observation. Then, we added Gaussian noise to each pixel on the projected maps of our mock observations with the rms given in Table~\ref{tab:1}.
One caveat of our study is that we did not simulate the contamination of the cosmic infrared background (CIB), which includes the clustering of dusty star-forming galaxies (DSFGs) and the shot noise from DSFGs and radio sources. At higher frequencies, the CIB becomes significantly contaminated by the SZ signal \citep{Hauser1998,Hauser2001}. The clustering of DSFGs can be computed by using the CIB power spectrum derived from CIB anisotropies surveys (e.g. \citealt{PlanckCollaboration2014}), which can then be extrapolated to the observation frequencies. For unresolved DSFGs and radio sources, the shot noise can be computed using the models from \citet{Bethermin2012} for DSFGs and \citet{Tucci2011} for radio sources. Additionally, contamination from compact radio sources and submillimeter point sources can, in principle, be subtracted using multiband and high-resolution observations. The CMB can also be modeled with the software \texttt{HEALPix} \citep{Gorski2005} to mitigate its contamination. Therefore, our analysis assumed that the CMB and CIB contaminations can be well separated.

\begin{figure*}
        \includegraphics[width=\textwidth]{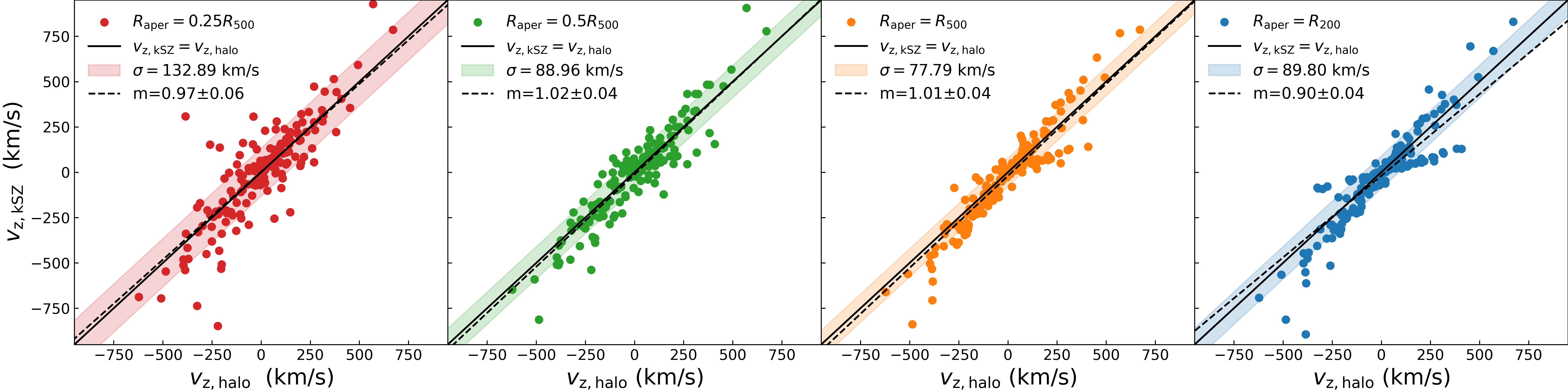}
    \caption{Comparison between the LOS (z-direction) velocities of galaxy clusters ($v_{\rm z,halo}$) given by TNG300-3 and the velocities obtained from the SZ effect ($v_{\rm z,kSZ}$) measured with different aperture sizes of $0.25R_{\rm 500}$, 0.5$R_{\rm 500}$, $R_{\rm 500}$, and $R_{\rm 200}$. The solid lines represent $v_{\rm z, halo} = v_{\rm z,kSZ}$, the colored regions represent the rms scatter of the $v_{\rm z, halo} = v_{\rm z,kSZ}$ line at a given x-axis for all galaxy clusters in our sample. The dashed lines are the best-fit lines of the data points excluding $|v_{\rm z,kSZ}|>500\ {\rm km\ s^{-1}}$, 
    and 'm' in the legends are the slopes of the best-fit lines.}
    \label{fig:vobs_vreal_nop0}
\end{figure*}

\begin{figure*}
        \includegraphics[width=\textwidth]{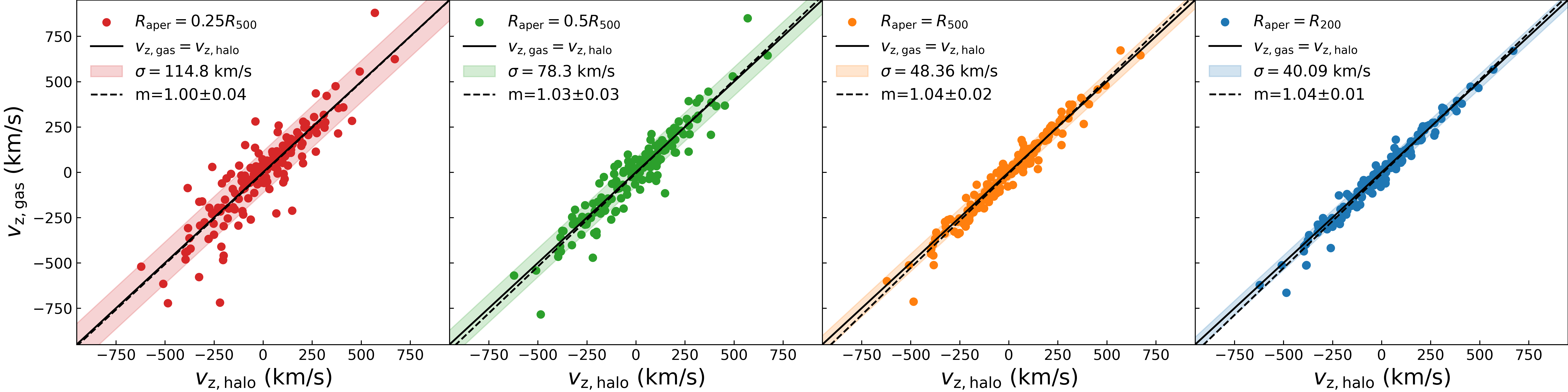}
    \caption{Comparison between mass-weighted gas velocities ($v_{\rm z,gas}$) and $v_{\rm z,halo}$ within different aperture sizes of 0.25$R_{\rm 500}$, 0.5$R_{\rm 500}$, $R_{\rm 500}$, and $R_{\rm 200}$. The solid lines represent $v_{\rm z,gas} = v_{\rm z,halo}$, the colored regions represent the rms scatter of the $v_{\rm z,gas} = v_{\rm z,halo}$ line at a given x-axis for all galaxy clusters in our sample. The dashed lines are the best-fit lines of the data points, and 'm' in the legends are the slopes of the best-fit lines.}
    \label{fig:vgas_vreal}
\end{figure*}

\section{Results}
\label{sec:results}

We projected the SZ signals of all gas particles onto the x-y plane for the galaxy clusters in our sample. An example of SZ effect maps at seven frequencies is shown in Figure~\ref{fig:SZmap}. After we obtained the projected maps of the SZ effect for each galaxy cluster, we measured the fluxes of the SZ effect within four different aperture sizes: 0.25$R_{\rm 500}$, 0.5$R_{\rm 500}$, $R_{\rm 500}$, and $R_{\rm 200}$ (corresponding to the circles in the SZ maps in Figure~\ref{fig:SZmap}). The four panels in Figure~\ref{fig:SZspectrum} depict examples of the derived broadband SED for these four apertures. We fit the data points with an analytic formula of the frequency for the SZ effect provided by \citet{Nozawa2006}, which describes the change in photon number density over time due to Compton scattering with electrons with relativistic corrections for a moving system. The fitting procedure allowed us to derive the best-fit results for the optical depth ($\tau$), gas temperatures ($kT$), and LOS (z-direction) peculiar velocities ($v_{\rm z,kSZ}$). 

The Compton $y$-parameter quantifies the photon energy change due to inverse-Compton scattering. The radial distribution of $y_{\rm tSZ}$ shows the distribution of the electron pressure within the cluster, which is typically highest at the center and decreases with increasing radius. We show the mean radial profiles of the tSZ effect ($y_{\rm tSZ}=\frac{k_{\rm B}T_{e}}{m_{e}c^{2}}\tau$) for the four mass bins of our sample in Figure~\ref{fig:ytSZ}. In all mass bins, the $y_{\rm tSZ}$ radial profiles decrease from the cluster center to the outer regions. More massive clusters typically exhibit higher $y_{\rm tSZ}$ radial profiles that correspond to a higher electron pressure.

Our primary focus was the $v_{\rm z,kSZ}$ estimated from the mock observations of the SZ effect. We first present the results without observational noise.
Figure~\ref{fig:vobs_vreal_nop0} presents one of the main results of our work, where we compare the $v_{\rm z,kSZ}$ and the true LOS (z-direction) velocities of the entire galaxy cluster ($v_{\rm z,halo}$) based on the measurements with various aperture sizes. The $v_{\rm z,kSZ}$ values were derived by fitting the SZ formula using a nonlinear least-squares function. As shown in the figure, there are outliers in all cases, and the slopes of the $v_{\rm z,kSZ}$-$v_{\rm z,halo}$ relations for all clusters in our sample are greater than one, ranging from 1.07 to 1.12. Galaxy clusters with a high measured $v_{\rm z,kSZ}$ are overestimated in LOS velocities and show large deviations from $v_{\rm z,halo}$, which is likely due to the discrepancy between $v_{\rm z,gas}$ and $v_{\rm z,halo}$, as shown in Figure~\ref{fig:vgas_vreal}. 
In Figure~\ref{fig:vgas_vreal}, we show the relations between the $v_{\rm z,gas}$ and $v_{\rm z,halo}$ for all galaxy clusters. The rms scatter decreases with larger measuring aperture sizes. The slopes of the $v_{\rm z,gas} - v_{\rm z,halo}$ relations of our selected clusters are slightly larger than one, which may also slightly contribute to the overestimation of the $v_{\rm z,kSZ}$, regardless of whether the galaxy clusters approach us or move away from us in Figure~\ref{fig:vobs_vreal_nop0}. 
We also checked the $v_{\rm z,gas} - v_{\rm z,halo}$ relations for $\sim$3000 TNG halos with higher redshift and mass ranges. The slopes are within $\sim 3\%$ of one.
These overestimates affect the slopes of the linear-fitting results, which are significantly larger than one. Thus, we fit the $v_{\rm z,kSZ}$-$v_{\rm z,halo}$ relation after excluding the clusters with $|v_{\rm z,kSZ}|>$ 500~$\rm km\ s^{-1}$. Except for the case with a photometric aperture size of $R_{\rm 200}$, the slopes are consistent with unity within the errors. There is a significant difference between the slope of the $v_{\rm z,kSZ}$-$v_{\rm z,halo}$ relation with photometric aperture sizes of $R_{\rm 200}$ (0.90) and unity, which is due to underestimates of $v_{\rm z,kSZ}$ for many clusters. The underestimates are likely caused by the large fitting errors with very low average fluxes of the SZ effect out to $R_{\rm 200}$. The root mean-square (rms) scatter around the $v_{\rm z,kSZ} = v_{\rm z, halo}$ line at the given x-axis can reach approximately 132.9~$\rm km\ s^{-1}$ for the smallest aperture size of 0.25$R_{\rm 500}$. The scatters roughly decrease with larger aperture sizes, and have values of $\sim$ 89.0~$\rm km\ s^{-1}$ and 77.8~$\rm km\ s^{-1}$ for apertures of 0.5$R_{\rm 500}$ and $R_{\rm 500}$, respectively. The rms scatter of the results obtained with apertures of $R_{\rm 200}$ increases slightly to $\sim$ 89.8~$\rm km\ s^{-1}$.

\begin{figure*}
        \includegraphics[width=\textwidth]{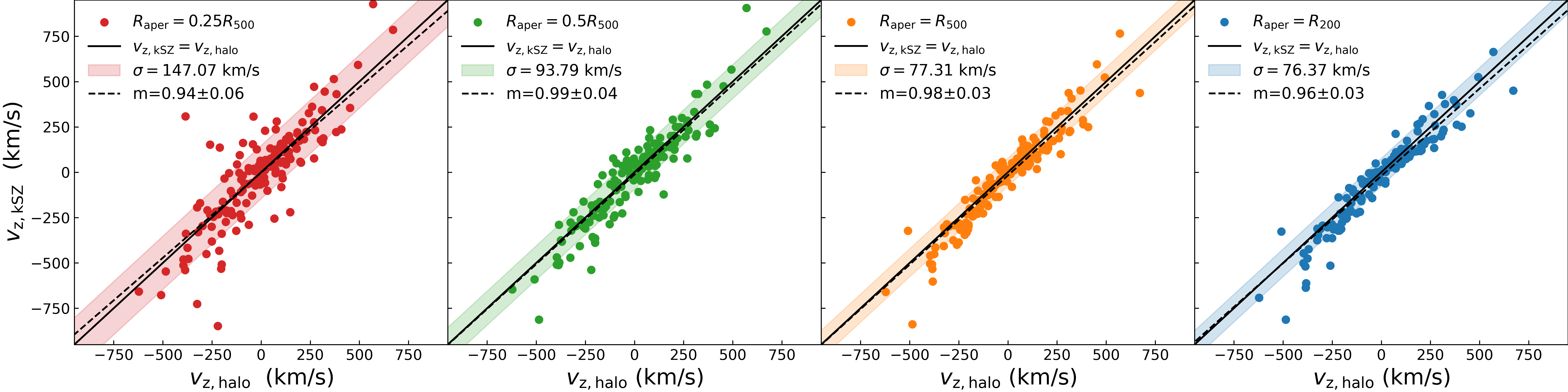}
    \caption{Comparison between the LOS (z-direction) velocities of the galaxy clusters ($v_{\rm z,halo}$) given by TNG300-3 and the velocities obtained with the SZ effect ($v_{\rm z,kSZ}$) measured with different aperture sizes of $0.25R_{\rm 500}$, 0.5$R_{\rm 500}$, $R_{\rm 500}$, and $R_{\rm 200}$, given the bounds for $kT$. The solid lines represent $v_{\rm z, halo} = v_{\rm z,kSZ}$, and the colored regions represent the rms scatter of the $v_{\rm z, halo} = v_{\rm z,kSZ}$ line at given x-axis for all galaxy clusters in our sample. The dashed lines are the best-fit lines of the data points excluding $|v_{\rm z,kSZ}|>500\ {\rm km\ s^{-1}}$, and 'm' in the legends are the slopes of the best-fit lines.}
    \label{fig:vobs_vreal_ex500_Tx_nop0}
\end{figure*}

We also tested the case with measurements of an X-ray spectroscopic temperature $T_{\rm x}$ by limiting $kT$ to be within 0.65$\sim$1.35 of the gas-mass-weighted temperature $T_{\rm gas}$. This bound for $kT$ is based on \citet{Nagai2007}, which studied the relation between $T_{\rm x}$ and $T_{\rm gas}$. Figure~\ref{fig:vobs_vreal_ex500_Tx_nop0} shows the $v_{\rm z,kSZ} - v_{\rm z, halo}$ relations with $v_{\rm z,kSZ}$ derived by fitting the SZ formula using the nonlinear least-squares function given the bounds of $kT$. We excluded the galaxy clusters with $|v_{\rm z,kSZ}|>$ 500~$\rm km\ s^{-1}$ to carry out the linear fitting to the $v_{\rm z,kSZ}$-$v_{\rm z,halo}$ relation. In this case, the rms scatter of the $v_{\rm z,kSZ} - v_{\rm z, halo}$ relations is smaller than the results in Figure~\ref{fig:vobs_vreal_nop0}, and the number of outliers decreases as well. This suggests that measurements of $T_{\rm x}$ are helpful when  $v_{\rm z,kSZ}$ is estimated from the kSZ effect, especially for galaxy clusters with faint SZ signals. The slopes, ranging from 0.94 to 0.99, are consistent with unity, which is also within the errors of the slopes. Our results differ from those of \citet{Wang2021}, who identified slopes much smaller than unity. They estimated the $v_{\rm z,kSZ}$ from temperature distortion maps of the SZ effect instead of simulated 2D observations and the derived broad SED at different frequencies as we did.

\begin{figure}
	\includegraphics[width=0.9\columnwidth]{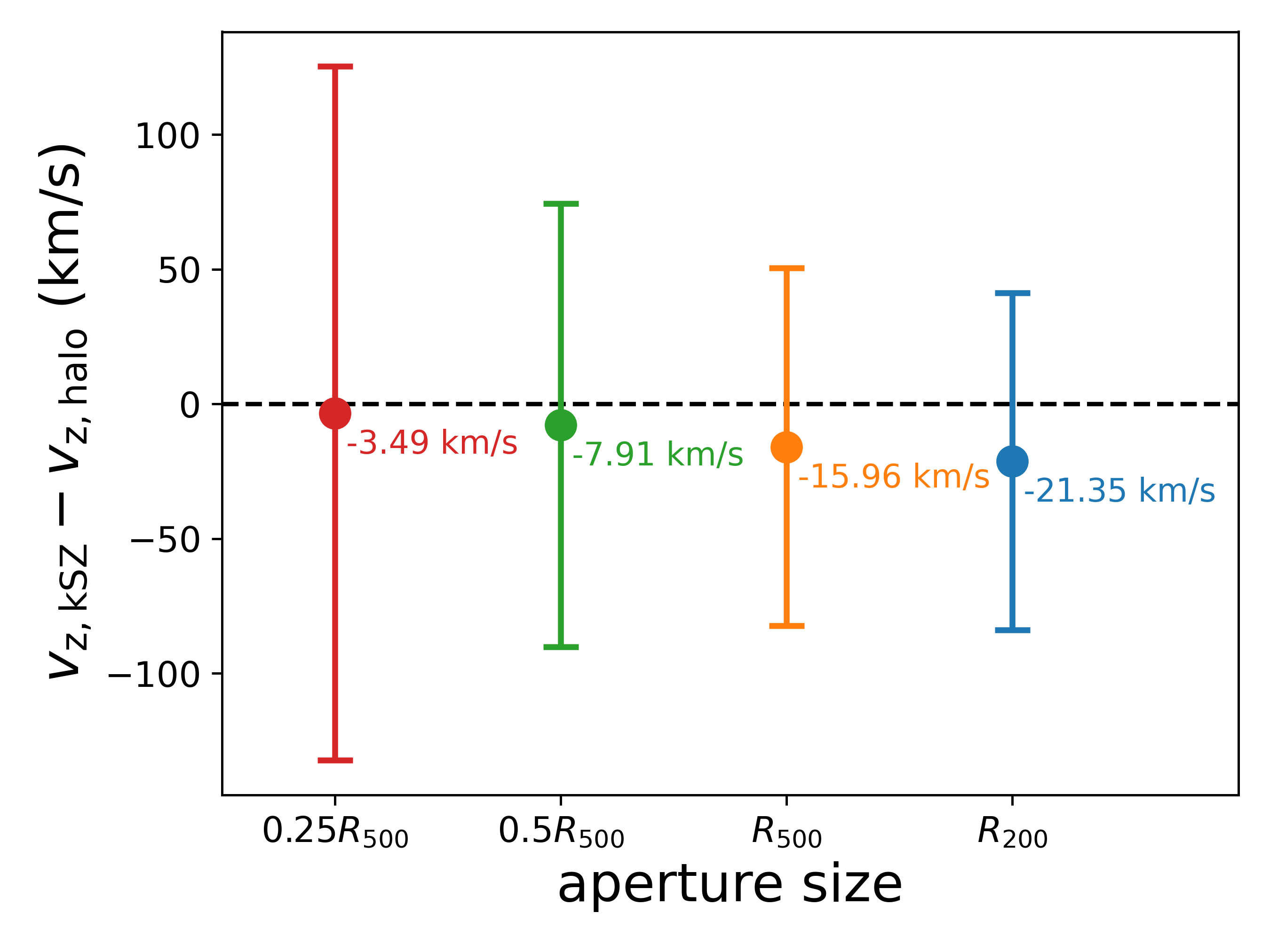}
    \caption{Median values (dots with values) of the difference between $v_{\rm z,kSZ}$ and $v_{\rm z,halo}$ with measurements using different aperture sizes. The error bars show the standard deviation of ($v_{\rm z,kSZ} - v_{\rm z,halo}$).}
    \label{fig:diff_v}
\end{figure}

\begin{figure*}
        \includegraphics[width=\textwidth]{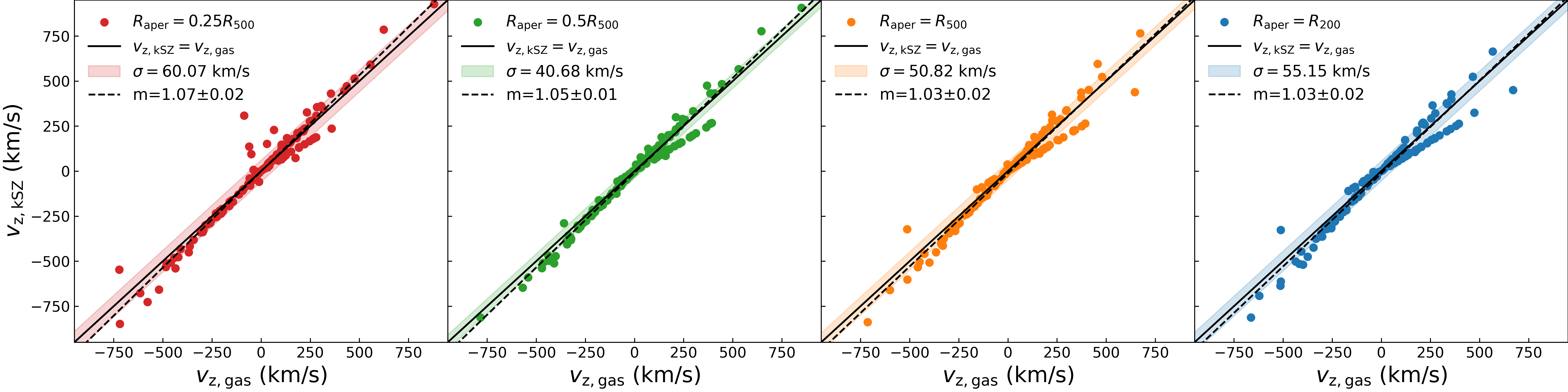}
    \caption{Comparison between $v_{\rm z,gas}$ and $v_{\rm z,kSZ}$ measured with different aperture sizes of 0.25$R_{\rm 500}$, 0.5$R_{\rm 500}$, $R_{\rm 500}$, and $R_{\rm 200}$, given the bounds for $kT$. The solid lines represent $v_{\rm z, kSZ} = v_{\rm z,gas}$, and the colored regions represent the rms scatter of the $v_{\rm z, kSZ} = v_{\rm z,gas}$ line at given x-axis for all galaxy clusters in our sample. The dashed lines are the best-fit lines of the data points, and 'm' in the legends are the slopes of the best-fit lines.}
    \label{fig:vobs_vgas}
\end{figure*}

In Figure~\ref{fig:diff_v}, we present the median values of the differences between $v_{\rm z,kSZ}$ and $v_{\rm z,halo}$, denoted as $v_{\rm z,kSZ} - v_{\rm z,halo}$, for the results obtained with different apertures under given bounds for $kT$. The error bars represent the standard deviations from the median of $v_{\rm z,kSZ} - v_{\rm z,halo}$. The median values of $v_{\rm z,kSZ} - v_{\rm z,halo}$ with different aperture sizes are all negative, which is possibly because it is easier to detect galaxy clusters that approach us. When galaxy clusters approach us, the flux of the kSZ effect is positive and increases the peak value of the SZ effect, which is easier to detect. The median of $v_{\rm z,kSZ} - v_{\rm z,halo}$ with larger apertures show larger deviations from 0~$\rm km\ s^{-1}$. We also compared the $v_{\rm z,kSZ}$ with the true velocity of hot gas -- $v_{\rm z,gas}$ in Figure~\ref{fig:vobs_vgas} and linearly fit the $v_{\rm z,kSZ}-v_{\rm z,gas}$ relations for all galaxy clusters. The slopes are all larger than one, indicating that in general, the method that uses the analytic formula fitting \citep{Nozawa2006} tends to overestimate the LOS peculiar velocities based on the kSZ effect, regardless of whether galaxy clusters approach us or move away from us.

\begin{figure*}
        \includegraphics[width=\textwidth]{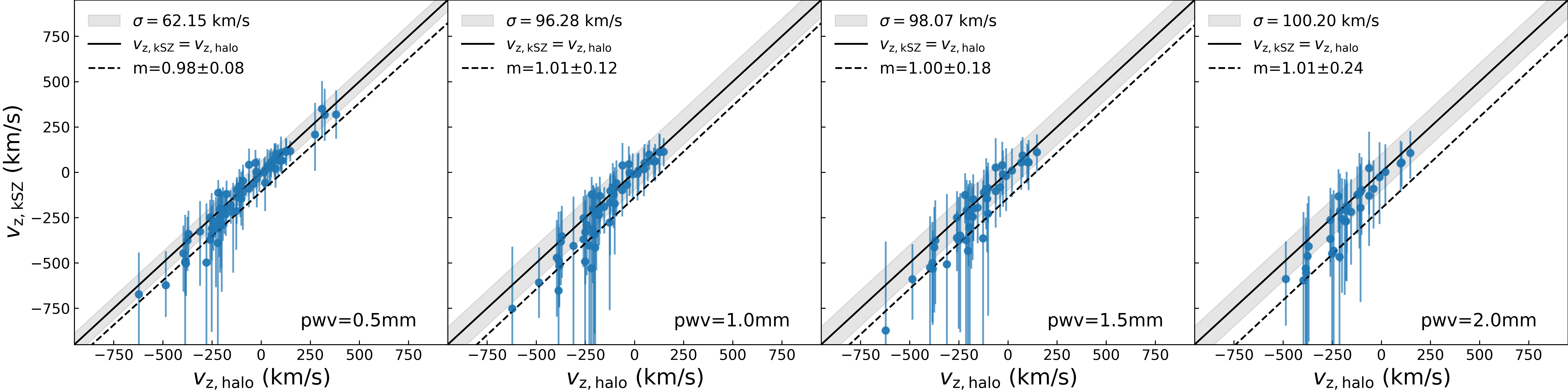}
    \caption{Comparison between $v_{\rm z,halo}$ and $v_{\rm z,kSZ}$ under different weather conditions (pwv = 0.5~mm, 1.0~mm, 1.5~mm, and 2.0~mm) with given bounds for $kT$ for a 30-meter radio telescope. The results in these four figures are based on measurements with an aperture size of $R_{\rm 200}$. The solid lines represent $v_{\rm z, kSZ} = v_{\rm z,halo}$, and the gray regions represent the rms scatters of the $v_{\rm z, kSZ} = v_{\rm z,halo}$ lines at a given x-axis for all galaxy clusters in our sample. The dashed lines are the best-fit lines of the data points excluding $|v_{\rm z,kSZ}|>500\ {\rm km\ s^{-1}}$, and 'm' in the legends are the slopes of the best-fit lines.}
    \label{fig:rms_pwv}
\end{figure*}

\begin{table*}
    \caption{$v_{\rm z, kSZ} - v_{\rm z,halo}$ relations for telescopes with different sizes under different PWV conditions.}
    \label{tab:2}
    \resizebox{1.0\textwidth}{!}{\begin{tabular}{cccccccccccccc}
    \hline
    Diameter & PWV & 
    \multicolumn{3}{c}{0.25$R_{500}$} & \multicolumn{3}{c}{0.50$R_{500}$} & \multicolumn{3}{c}{$R_{500}$} & \multicolumn{3}{c}{$R_{200}$}\\
    \cline{3-14}
     (m) & (mm) & slope & $\sigma_{\rm rms}$ & $f_{\rm detect}$ & slope & $\sigma_{\rm rms}$ & $f_{\rm detect}$ & slope & $\sigma_{\rm rms}$ & $f_{\rm detect}$ & slope & $\sigma_{\rm rms}$ & $f_{\rm detect}$ \\
    \hline
   10 & 0.5 & 0.89$\pm$0.12 & 154.6 & 32$\%$ & 0.85$\pm$0.16 & 119.5 & 25$\%$ & 0.92$\pm$0.42 & 119.6 & 13$\%$ & - & - & 3$\%$ \\
       & 1.0 & 0.87$\pm$0.16 & 156.9 & 25$\%$ & 0.91$\pm$0.26 & 154.4 & 19$\%$ & - & - & 5$\%$ & - & - & 1$\%$ \\
       & 1.5 & 0.87$\pm$0.26 & 226.7 & 15$\%$ & 0.88$\pm$0.46 & 125.6 & 11$\%$ & - & - & 3$\%$ & - & - & 0$\%$ \\
       & 2.0 & 0.78$\pm$0.43 & 169.2 & 12$\%$ & 1.24$\pm$0.98 & 141.9 & 7$\%$ & - & - & 1$\%$ & - & - & 0$\%$ \\
    \hline
    
    30 & 0.5 & 0.93$\pm$0.06 & 129.6 & 82$\%$ & 0.94$\pm$0.05 & 92.1 & 78$\%$ & 0.91$\pm$0.05 & 75.8 & 62$\%$ & 0.98$\pm$0.08 & 62.2 & 46$\%$ \\
       & 1.0 & 0.92$\pm$0.06 & 139.3 & 75$\%$ & 0.90$\pm$0.05 & 102.9 & 67$\%$ & 0.91$\pm$0.07 & 63.4 & 48$\%$ &1.01$\pm$0.12 & 96.3 & 36$\%$ \\
       & 1.5 & 0.90$\pm$0.07 & 142.1 & 65$\%$ & 0.85$\pm$0.06 & 99.3 & 56$\%$ & 0.92$\pm$0.09 & 72.6 & 45$\%$ & 1.00$\pm$0.18 & 98.1 & 28$\%$ \\
       & 2.0 & 0.85$\pm$0.07 & 145.2 & 59$\%$ & 0.81$\pm$0.07 & 106.5 & 51$\%$ & 0.94$\pm$0.11 & 89.0 & 40$\%$ & 1.01$\pm$0.24 & 100.2 & 21$\%$ \\
    \hline
    
    100 & 0.5 & 0.87$\pm$0.06 & 138.1 & 98$\%$ & 1.00$\pm$0.04 & 89.8 & 94$\%$ & 1.05$\pm$0.03 & 62.4 & 90$\%$ & 1.01$\pm$0.03 & 51.0 & 86$\%$ \\
        & 1.0 & 0.89$\pm$0.06 & 139.8 & 90$\%$ & 0.98$\pm$0.04 & 90.7 & 88$\%$ & 1.01$\pm$0.03 & 59.3 & 85$\%$ & 0.97$\pm$0.03 & 50.6 & 78$\%$ \\
        & 1.5 & 0.88$\pm$0.06 & 137.7 & 88$\%$ & 0.98$\pm$0.04 & 90.0 & 86$\%$ & 0.98$\pm$0.03 & 57.4 & 82$\%$ & 0.95$\pm$0.03 & 53.8 & 72$\%$ \\
        & 2.0 & 0.89$\pm$0.06 & 137.5 & 86$\%$ & 0.96$\pm$0.04 & 89.6 & 85$\%$ & 0.96$\pm$0.04 & 57.9 & 75$\%$ & 0.94$\pm$0.04 & 54.9 & 65$\%$ \\
    \hline
    \end{tabular}}
    \begin{tablenotes}
      \small
      \item Notes: For cases for which too few sources were detected for at least three frequencies, we cannot fit the relation between $v_{\rm z, kSZ}$ and $v_{\rm z,halo}$. These cases are labeled with a minus. The rms scatters, $\sigma_{\rm rms}$, are in km/s. The $f_{\rm detect}$ is the detection rate of galaxy clusters that have measurements with an S/N greater than 2 for at least three frequencies.
    \end{tablenotes}
\end{table*}

We further accounted for observational noises when simulating observations of the SZ effect. We adopted the noises observed with 10-meter, 30-meter, and 100-meter telescopes with an elevation of 45 degrees operating under various PWV conditions at the CBI as listed in Table~\ref{tab:1}. We conducted the measurements with aperture sizes of $0.25R_{\rm 500}$, 0.5$R_{\rm 500}$, $R_{\rm 500}$, and $R_{\rm 200}$ separately. The results for the 30-meter telescope, with an aperture size of $R_{\rm 200}$ for the SZ flux measurement, are visualized in Figure~\ref{fig:rms_pwv}. We only included clusters with measurements with a signal-to-noise ratio (S/N) greater than 2 for at least three frequencies. The $v_{\rm z,kSZ}$ values were derived by fitting the SZ signal with observational errors using a Markov chain Monte Carlo (MCMC) method. The linear fitting results of the $v_{\rm z,kSZ} - v_{\rm z,halo}$ relation excluding the clusters with $|v_{\rm z,kSZ}|>$ 500~$\rm km\ s^{-1}$ for all cases are listed in Table~\ref{tab:2}. As shown in Table~\ref{tab:2}, except for the 100-meter telescope, fewer sources can be detected with increasing PWV, and in general, the slope errors also increase slightly. 
For the 100-meter telescope, a decrease in PWV does not significantly improve the sensitivity and the accuracy of $v_{\rm z,kSZ}$. Instead, additional clusters with weak SZ signals will be detected under lower PWV, which introduces more uncertainties.
Overall, in most cases, the slopes are consistent with one within the errors. However, the rms scatter of the $v_{\rm z,kSZ}=v_{\rm z,halo}$ line at the given x-axis does not exhibit a clear decreasing trend with reduced observational noises. With lower observation noise, more faint sources are detected, but their average SZ fluxes are very low, leading to larger errors in estimating $v_{\rm z,kSZ}$ and larger rms scatter of the $v_{\rm z,kSZ}=v_{\rm z,halo}$ line at the given x-axis.

\section{Discussion}
\label{sec:discussion}

\begin{figure}
	\includegraphics[width=\columnwidth]{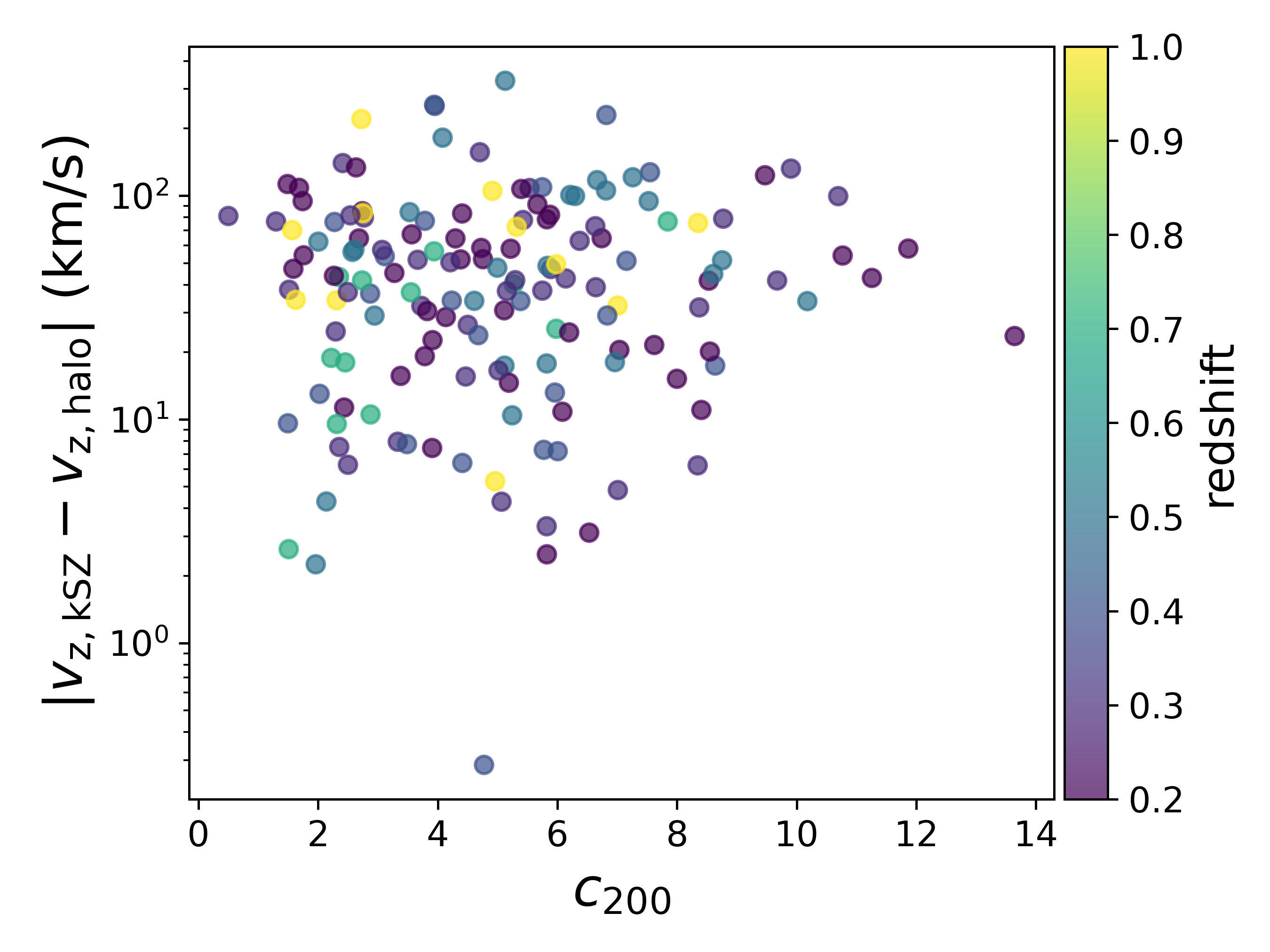}
    \caption{Distribution of $|v_{\rm z, kSZ} - v_{\rm z,halo}|$ against the concentration ($c_{\rm 200}$) of galaxy clusters in our sample for measurements with $R_{\rm aper}=R_{\rm 200}$. The colors of the data points represent their redshift.}
    \label{fig:c200}
\end{figure}

Any departure from spherical symmetry in the mass distribution of a galaxy cluster can introduce bias into the estimation of key physical properties such as mass, centroid, and peculiar velocity \citep{Lee2018,Yuan2020}. Consequently, a disturbed structure can have a significant impact on the cosmological constraints derived from various relations, including the $Y_{\rm SZ}-M$ relation and the concentration-mass relation \citep{Maccio2007,Ludlow2012}. The bias in estimating physical properties from SZ effect images depends on the dynamical state of the galaxy clusters \citep{Yuan2020}.

\begin{figure}
	\includegraphics[width=\columnwidth]{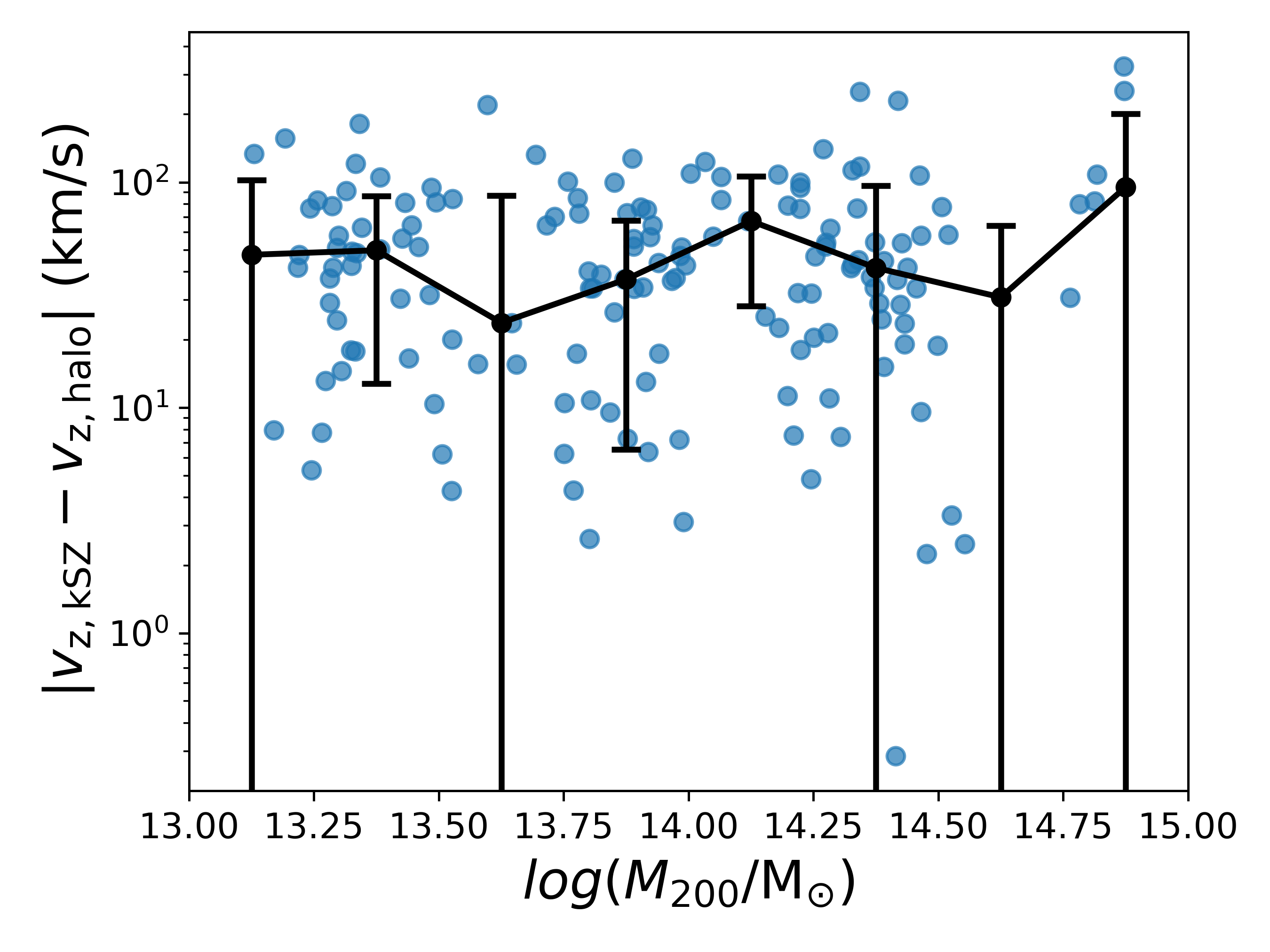}
    \caption{Distribution of $|v_{\rm z, kSZ} - v_{\rm z,halo}|$ against $\rm M_{200}$ of galaxy clusters in our sample for measurements with $\rm R_{aper}=R_{200}$. The black dots with error bars are the medians of $|v_{\rm z, kSZ} - v_{\rm z,halo}|$ and their standard deviations at different $M_{\rm 200}$.
    }
    \label{fig:M200}
\end{figure}

The concentration parameter, $c_{\rm 200}$, represents the central density of the halo and is used to characterize the assembly history, which reflects the time when it formed \citep{Bullock2001,Eke2001,Zhao2003}. We compare $|v_{\rm z, kSZ} - v_{\rm z,halo}|$ and $c_{\rm 200}$ in Figure~\ref{fig:c200}. In TNG300-3 run, the DM particles have the same mass of $\sim 0.255\times10^{10} {\rm M_{\odot}}/h$ given in MassTable in \texttt{Header} field, and the positions of every DM particle are given in \texttt{PartType1} snapshot fields of the TNG dataset. We fit the DM mass density distribution with the Navarro–Frenk–White (NFW) profile \citep{Navarro1996,Navarro1997},
\begin{equation}
    \rho_{\rm NFW}(r) = \frac{\rho_{\rm s}}{(r/r_{\rm s})(1+r/r_{\rm s})^2}.
\end{equation}
Then, we obtained $r_{\rm s}$ by fitting the NFW formula to the DM density profile. $R_{200}$ is the radius within which the mean density of the halo is 200 times the critical density, and it can be derived based on the radial density profile of all particles (DM, gas, stars, and black holes). The concentration, $c_{\rm 200}$ is $r_{\rm 200}/r_{\rm s}$. We do not observe an obvious trend between $|v_{\rm z, kSZ} - v_{\rm z,halo}|$ and $c_{\rm 200}$. This differs from our expectation that higher $c_{\rm 200}$ would lead to higher $|v_{\rm z, kSZ} - v_{\rm z,halo}|$ because clusters showing signs of dynamical activity tend to exhibit high concentrations \citep{Sereno2013}, and there exists an empirical relation $M_{\rm 200}-c_{\rm 200}$ within galaxy clusters, where the concentration decreases as the halo mass increases \citep{Buote2007,Comerford2007,Ludlow2012,Du2015,Amodeo2016}.  In Figure~\ref{fig:M200}, we compare $M_{\rm 200}$ and $|v_{\rm z, kSZ} - v_{\rm z,halo}|$ and find no obvious trend in the median value of $|v_{\rm z, kSZ} - v_{\rm z,halo}|$ in the halo mass bins either.

\begin{figure}
	\includegraphics[width=\columnwidth]{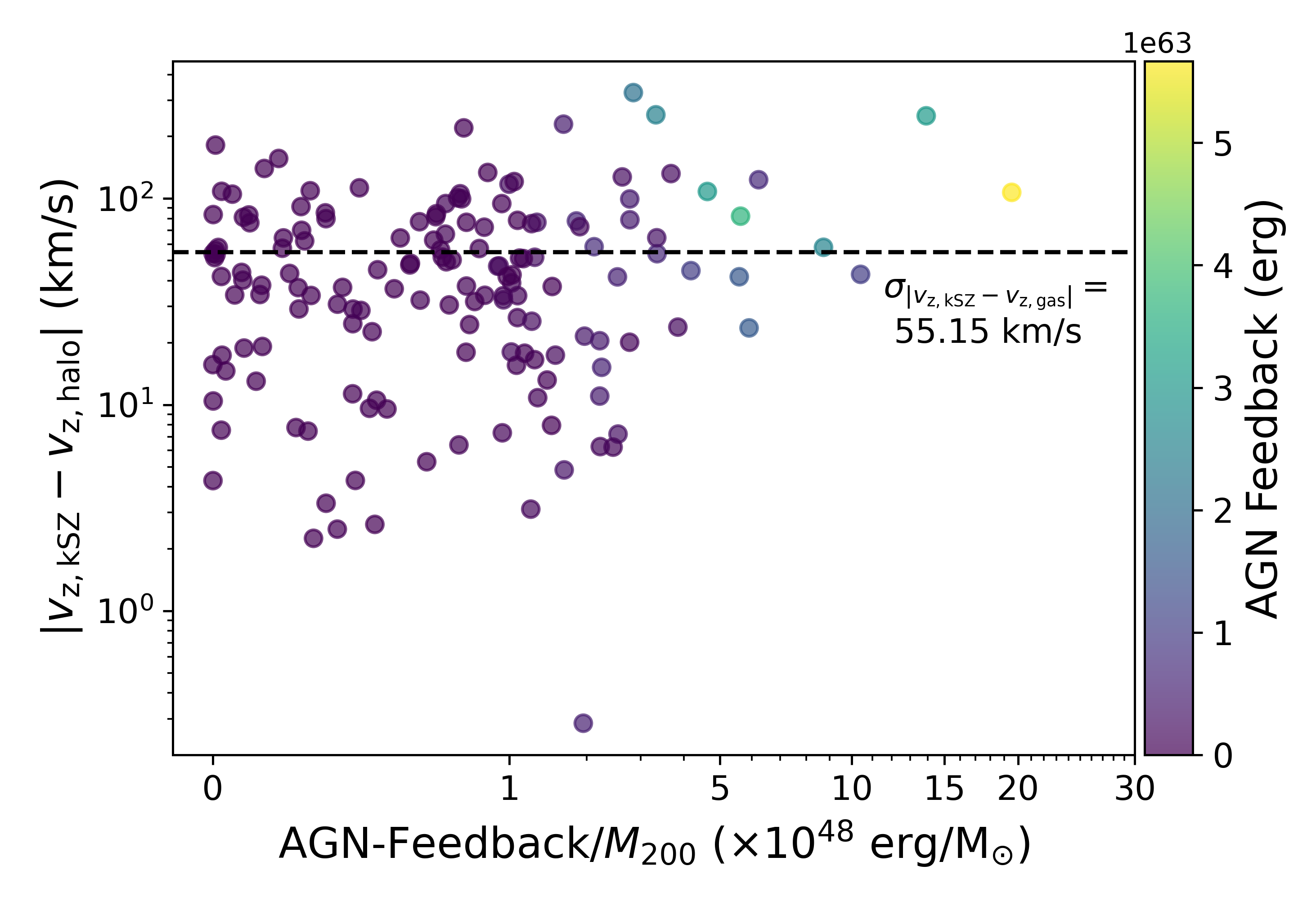}
    \caption{Distribution of $|v_{\rm z, kSZ} - v_{\rm z,halo}|$ against the ratios of total cumulative AGN feedback to $\rm M_{200}$ of galaxy clusters in our sample for measurements with $\rm R_{aper}=R_{200}$. The dashed line represents the rms scatter ($\sim$55.15~$\rm km\ s^{-1}$) of the $v_{\rm z,kSZ} = v_{\rm z,gas}$ line at a given x-axis for all galaxy clusters based on measurements using apertures of $R_{\rm 200}$, as shown in Figure~\ref{fig:vobs_vgas}. The color bar represents the AGN feedback of the galaxy clusters.}
    \label{fig:AGN}
\end{figure}

\begin{figure}
    \includegraphics[width=0.9\columnwidth]{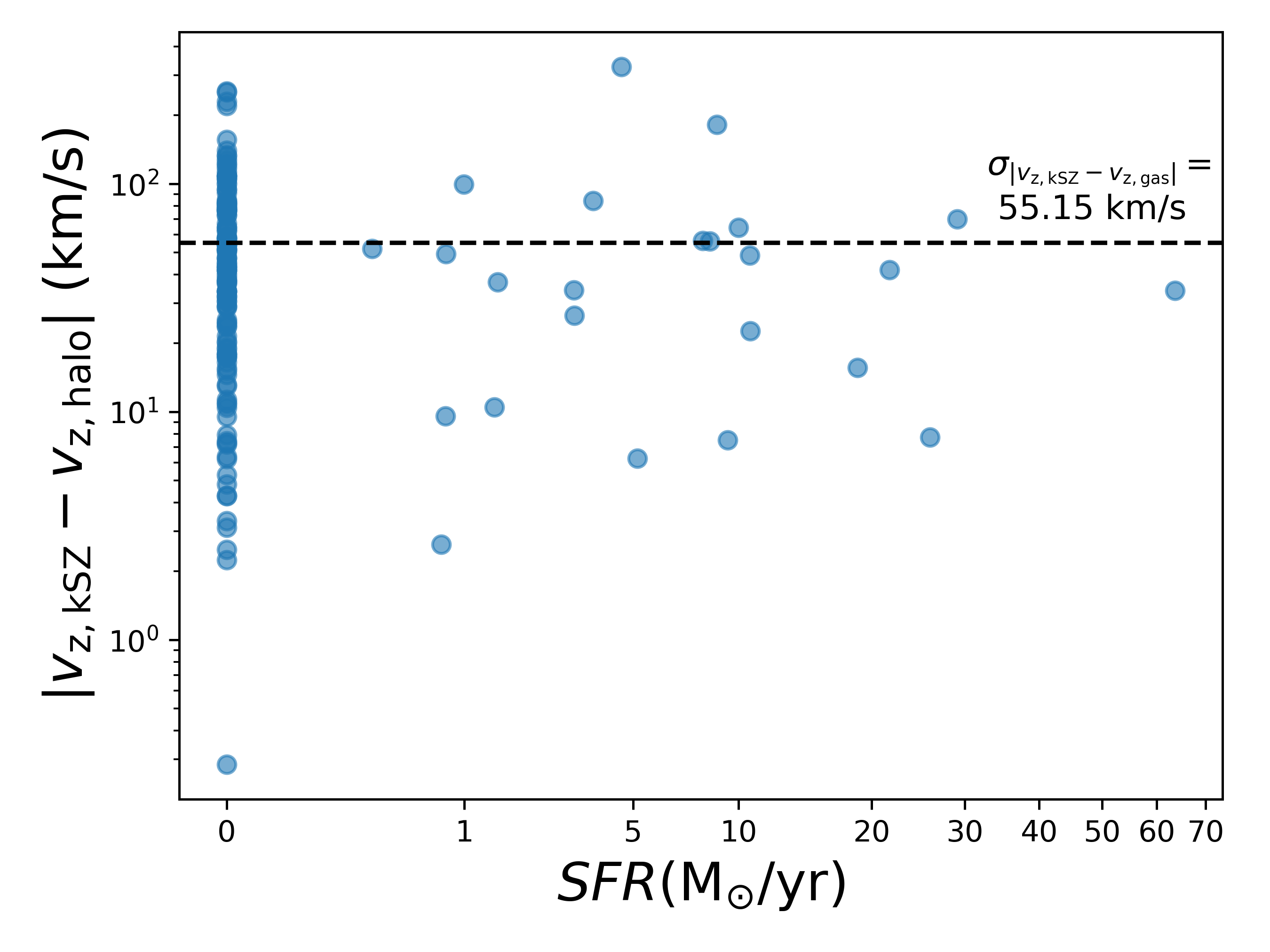}
    \caption{Distribution of $|v_{\rm z, kSZ} - v_{\rm z,halo}|$ against the SFR of galaxy clusters in our sample for measurements with $\rm R_{aper}=R_{200}$. The dashed line represents the rms scatter ($\sim$55.15~$\rm km\ s^{-1}$) of the $v_{\rm z,kSZ} = v_{\rm z,halo}$ line at a given x-axis for all galaxy clusters based on measurements using apertures of $R_{\rm 200}$, as shown in Figure~\ref{fig:vobs_vgas}.}
    \label{fig:stellar}
\end{figure}

Physical processes such as AGN and star formation activity can affect the dynamical state of galaxy clusters. We plot the AGN-feedback/$M_{\rm 200}-|v_{\rm z, kSZ} - v_{\rm z,halo}|$ relation in Figure~\ref{fig:AGN}. In this figure, most clusters with high AGN-feedback/$M_{\rm 200}$ exhibit relatively high $|v_{\rm z, kSZ} - v_{\rm z,halo}|$, which can reach tens to some hundred $\rm km\ s^{-1}$. The dashed line in Figure~\ref{fig:AGN} represents the rms scatter ($\sim$ 55.15~$\rm km~s^{-1}$) of the $v_{\rm z,kSZ} = v_{\rm z,gas}$ line at a given x-axis with a photometric aperture size of $R_{\rm 200}$, which reflects the error introduced by fitting the SZ spectrum with an analytic formula. This suggests that AGN feedback indeed induces deviations from equilibrium in galaxy clusters, and the perturbation of the dynamical state of clusters caused by AGN feedback appears to outweigh the influence of the self-gravitational potential energy of the cluster. Most of the clusters with high AGN feedback are located around or largely above the value of the fitting error. 
Although $|v_{\rm z, kSZ} - v_{\rm z,halo}|$ for clusters with strong AGN feedback are relatively high and may reflect the effect of AGN feedback on the cluster dynamical state, these velocity differences are likely to be obscured for measurement errors and contamination. Figure~\ref{fig:stellar} shows the relation between the star formation rate (SFR) and $|v_{\rm z, kSZ} - v_{\rm z,halo}|$ relation. Clusters with high SFR show a relatively high $|v_{\rm z, kSZ} - v_{\rm z,halo}|$, but their $|v_{\rm z, kSZ} - v_{\rm z,halo}|$ are just about or lower than the rms scatter of the $v_{\rm z,kSZ} = v_{\rm z,gas}$ line at the given x-axis.

\section{Conclusions}
\label{sec:conclusion}

We selected a sample of galaxy clusters with varying redshift and total mass from TNG300-3 and generated mock 2D maps of the SZ effect at seven frequencies. We measured the fluxes of the SZ effect and fit the spectra with an analytic formula \citep{Nozawa2006}, and we obtained the best-fit parameters including the peculiar velocity $v_{\rm z,kSZ}$. We compared the velocity $v_{\rm z,kSZ}$  with the true LOS peculiar velocities $v_{\rm z,halo}$ of the halos. The main conclusions are listed below.

(1) Overall, $v_{\rm z,kSZ}$ represents the halo velocity well. For
high peculiar velocities, they tend to be overestimated. 
When they are excluded, the slopes of the $v_{\rm z, kSZ} - v_{\rm z,halo}$ relations are consistent with unity within the errors.

(2) The median values of ($v_{\rm z, kSZ} - v_{\rm z,halo}$) are lower than 0~$\rm km\ s^{-1}$, indicating that it is easier to detect the kSZ effect of galaxy clusters that approach us.

(3) When we considered the observational noise, most of the slopes of $v_{\rm z, kSZ} - v_{\rm z,halo}$ relations are consistent with one within the errors. In general, with lower observation noise, more sources can be detected, and the slope errors decrease slightly.

(4) The $|v_{\rm z, kSZ} - v_{\rm z,halo}|$ exhibits no discernible trend with increasing $c_{\rm 200}$ parameter, which is associated with the halo assembly history. In addition, the $|v_{\rm z, kSZ} - v_{\rm z,halo}|$ shows no obvious trend with increasing $M_{\rm 200}$.

(5) Physical processes such as AGN feedback and star formation can influence the dynamical states of hot gas in galaxy clusters. The clusters with high ${\rm AGN\ feedback}/M_{\rm 200}$ and SFR exhibit relatively large discrepancies between $v_{\rm z, kSZ}$ and $v_{\rm z,halo}$ that reach several tens to some hundred $\rm km\ s^{-1}$. These variations may influence the peculiar velocity estimates derived with the kSZ effect. However, many of these values are comparable to or lower than the average fitting error.
 
\begin{acknowledgements} 
This work acknowledges the support from the National Key R\&D Program of China No. 2023YFA1608204, the National Natural Science Foundation of China (NSFC grants 12141301, 12121003, 12333002).
\end{acknowledgements}





\bibliographystyle{aa-yang}
\bibliography{SZE.bib}

\begin{thebibliography}{62}
\expandafter\ifx\csname natexlab\endcsname\relax\def\natexlab#1{#1}\fi

\bibitem[{{Adam} {et~al.}(2018){Adam}, {Adane}, {Ade}, {Andr{\'e}},
  {Andrianasolo}, {Aussel}, {Beelen}, {Beno{\^\i}t}, {Bideaud}, {Billot},
  {Bourrion}, {Bracco}, {Calvo}, {Catalano}, {Coiffard}, {Comis}, {De Petris},
  {D{\'e}sert}, {Doyle}, {Driessen}, {Evans}, {Goupy}, {Kramer}, {Lagache},
  {Leclercq}, {Leggeri}, {Lestrade}, {Mac{\'\i}as-P{\'e}rez}, {Mauskopf},
  {Mayet}, {Maury}, {Monfardini}, {Navarro}, {Pascale}, {Perotto}, {Pisano},
  {Ponthieu}, {Rev{\'e}ret}, {Rigby}, {Ritacco}, {Romero}, {Roussel}, {Ruppin},
  {Schuster}, {Sievers}, {Triqueneaux}, {Tucker}, \& {Zylka}}]{Adam2018}
{Adam}, R., {Adane}, A., {Ade}, P.~A.~R., {et~al.} 2018,
  \href{http://dx.doi.org/10.1051/0004-6361/201731503}{\color{blue}\aap},
  \href{https://ui.adsabs.harvard.edu/abs/2018A&A...609A.115A}{609, A115}

\bibitem[{{Adam} {et~al.}(2017){Adam}, {Bartalucci}, {Pratt}, {Ade},
  {Andr{\'e}}, {Arnaud}, {Beelen}, {Beno{\^\i}t}, {Bideaud}, {Billot},
  {Bourdin}, {Bourrion}, {Calvo}, {Catalano}, {Coiffard}, {Comis}, {D'Addabbo},
  {De Petris}, {D{\'e}mocl{\`e}s}, {D{\'e}sert}, {Doyle}, {Egami}, {Ferrari},
  {Goupy}, {Kramer}, {Lagache}, {Leclercq}, {Mac{\'\i}as-P{\'e}rez},
  {Maurogordato}, {Mauskopf}, {Mayet}, {Monfardini}, {Mroczkowski}, {Pajot},
  {Pascale}, {Perotto}, {Pisano}, {Pointecouteau}, {Ponthieu}, {Rev{\'e}ret},
  {Ritacco}, {Rodriguez}, {Romero}, {Ruppin}, {Schuster}, {Sievers},
  {Triqueneaux}, {Tucker}, {Zemcov}, \& {Zylka}}]{Adam2017}
{Adam}, R., {Bartalucci}, I., {Pratt}, G.~W., {et~al.} 2017,
  \href{http://dx.doi.org/10.1051/0004-6361/201629182}{\color{blue}\aap},
  \href{https://ui.adsabs.harvard.edu/abs/2017A&A...598A.115A}{598, A115}

\bibitem[{{Adams} \& {Blake}(2017)}]{Adams2017}
{Adams}, C. \& {Blake}, C. 2017,
  \href{http://dx.doi.org/10.1093/mnras/stx1529}{\color{blue}\mnras},
  \href{https://ui.adsabs.harvard.edu/abs/2017MNRAS.471..839A}{471, 839}

\bibitem[{{Adams} \& {Blake}(2020)}]{Adams2020}
{Adams}, C. \& {Blake}, C. 2020,
  \href{http://dx.doi.org/10.1093/mnras/staa845}{\color{blue}\mnras},
  \href{https://ui.adsabs.harvard.edu/abs/2020MNRAS.494.3275A}{494, 3275}

\bibitem[{{Allen} {et~al.}(2011){Allen}, {Evrard}, \& {Mantz}}]{Allen2011}
{Allen}, S.~W., {Evrard}, A.~E., \& {Mantz}, A.~B. 2011,
  \href{http://dx.doi.org/10.1146/annurev-astro-081710-102514}{\color{blue}\araa},
  \href{https://ui.adsabs.harvard.edu/abs/2011ARA&A..49..409A}{49, 409}

\bibitem[{{Alonso} {et~al.}(2016){Alonso}, {Louis}, {Bull}, \&
  {Ferreira}}]{Alonso2016}
{Alonso}, D., {Louis}, T., {Bull}, P., \& {Ferreira}, P.~G. 2016,
  \href{http://dx.doi.org/10.1103/PhysRevD.94.043522}{\color{blue}\prd},
  \href{https://ui.adsabs.harvard.edu/abs/2016PhRvD..94d3522A}{94, 043522}

\bibitem[{{Amodeo} {et~al.}(2016){Amodeo}, {Ettori}, {Capasso}, \&
  {Sereno}}]{Amodeo2016}
{Amodeo}, S., {Ettori}, S., {Capasso}, R., \& {Sereno}, M. 2016,
  \href{http://dx.doi.org/10.1051/0004-6361/201527630}{\color{blue}\aap},
  \href{https://ui.adsabs.harvard.edu/abs/2016A&A...590A.126A}{590, A126}

\bibitem[{{B{\'e}thermin} {et~al.}(2012){B{\'e}thermin}, {Daddi}, {Magdis},
  {Sargent}, {Hezaveh}, {Elbaz}, {Le Borgne}, {Mullaney}, {Pannella}, {Buat},
  {Charmandaris}, {Lagache}, \& {Scott}}]{Bethermin2012}
{B{\'e}thermin}, M., {Daddi}, E., {Magdis}, G., {et~al.} 2012,
  \href{http://dx.doi.org/10.1088/2041-8205/757/2/L23}{\color{blue}\apjl},
  \href{https://ui.adsabs.harvard.edu/abs/2012ApJ...757L..23B}{757, L23}

\bibitem[{{Birkinshaw}(1999)}]{Birkinshaw1999}
{Birkinshaw}, M. 1999,
  \href{http://dx.doi.org/10.1016/S0370-1573(98)00080-5}{\color{blue}\physrep},
  \href{https://ui.adsabs.harvard.edu/abs/1999PhR...310...97B}{310, 97}

\bibitem[{{Bullock} {et~al.}(2001){Bullock}, {Kolatt}, {Sigad}, {Somerville},
  {Kravtsov}, {Klypin}, {Primack}, \& {Dekel}}]{Bullock2001}
{Bullock}, J.~S., {Kolatt}, T.~S., {Sigad}, Y., {et~al.} 2001,
  \href{http://dx.doi.org/10.1046/j.1365-8711.2001.04068.x}{\color{blue}\mnras},
  \href{https://ui.adsabs.harvard.edu/abs/2001MNRAS.321..559B}{321, 559}

\bibitem[{{Buote} {et~al.}(2007){Buote}, {Gastaldello}, {Humphrey},
  {Zappacosta}, {Bullock}, {Brighenti}, \& {Mathews}}]{Buote2007}
{Buote}, D.~A., {Gastaldello}, F., {Humphrey}, P.~J., {et~al.} 2007,
  \href{http://dx.doi.org/10.1086/518684}{\color{blue}\apj},
  \href{https://ui.adsabs.harvard.edu/abs/2007ApJ...664..123B}{664, 123}

\bibitem[{{Carlstrom} {et~al.}(2002){Carlstrom}, {Holder}, \&
  {Reese}}]{Carlstrom2002}
{Carlstrom}, J.~E., {Holder}, G.~P., \& {Reese}, E.~D. 2002,
  \href{http://dx.doi.org/10.1146/annurev.astro.40.060401.093803}{\color{blue}\araa},
  \href{https://ui.adsabs.harvard.edu/abs/2002ARA&A..40..643C}{40, 643}

\bibitem[{{Catalano} {et~al.}(2018){Catalano}, {Adam}, {Ade}, {Andr{\'e}},
  {Aussel}, {Beelen}, {Beno{\^\i}t}, {Bideaud}, {Billot}, {Bourrion}, {Calvo},
  {Comis}, {De Petris}, {D{\'e}sert}, {Doyle}, {Driessen}, {Goupy}, {Kramer},
  {Lagache}, {Leclercq}, {Lestrade}, {Mac{\'\i}as-P{\'e}rez}, {Mauskopf},
  {Mayet}, {Monfardini}, {Pascale}, {Perotto}, {Pisano}, {Ponthieu},
  {Rev{\'e}ret}, {Ritacco}, {Romero}, {Roussel}, {Ruppin}, {Schuster},
  {Sievers}, {Triqueneaux}, {Tucker}, {Zylka}, {Barria}, {Bres}, {Camus},
  {Chanthib}, {Donnier-Valentin}, {Exshaw}, {Garde}, {Gerardin}, {Leggeri},
  {Levy-Bertrand}, {Guttin}, {Hoarau}, {Grollier}, {Mocellin}, {Pont},
  {Rodenas}, {Tissot}, {Galvez}, {John}, {Ungerechts}, {Sanchez}, {Mellado},
  {Munoz}, {Pierfederici}, {Penalver}, {Navarro}, {Bosson}, {Bouly}, {Bouvier},
  {Geraci}, {Li}, {Menu}, {Ponchant}, {Roni}, {Roudier}, {Scordillis},
  {Tourres}, {Vescovi}, {Barbier}, {Billon-Pierron}, {Adane}, {Andrianasolo},
  {Bracco}, {Coiffard}, {Evans}, {Maury}, \& {Rigby}}]{Catalano2018}
{Catalano}, A., {Adam}, R., {Ade}, P.~A.~R., {et~al.} 2018,
  \href{http://dx.doi.org/10.1007/s10909-018-1884-5}{\color{blue}Journal of Low
  Temperature Physics},
  \href{https://ui.adsabs.harvard.edu/abs/2018JLTP..193..916C}{193, 916}

\bibitem[{{CCAT-Prime Collaboration} {et~al.}(2023){CCAT-Prime Collaboration},
  {Aravena}, {Austermann}, {Basu}, {Battaglia}, {Beringue}, {Bertoldi},
  {Bigiel}, {Bond}, {Breysse}, {Broughton}, {Bustos}, {Chapman}, {Charmetant},
  {Choi}, {Chung}, {Clark}, {Cothard}, {Crites}, {Dev}, {Douglas}, {Duell},
  {D{\"u}nner}, {Ebina}, {Erler}, {Fich}, {Fissel}, {Foreman}, {Freundt},
  {Gallardo}, {Gao}, {Garc{\'\i}a}, {Giovanelli}, {Golec}, {Groppi}, {Haynes},
  {Henke}, {Hensley}, {Herter}, {Higgins}, {Hlo{\v{z}}ek}, {Huber}, {Huber},
  {Hubmayr}, {Jackson}, {Johnstone}, {Karoumpis}, {Keating}, {Komatsu}, {Li},
  {Magnelli}, {Matthews}, {Mauskopf}, {McMahon}, {Meerburg}, {Meyers},
  {Muralidhara}, {Murray}, {Niemack}, {Nikola}, {Okada}, {Puddu}, {Riechers},
  {Rosolowsky}, {Rossi}, {Rotermund}, {Roy}, {Sadavoy}, {Schaaf}, {Schilke},
  {Scott}, {Simon}, {Sinclair}, {Sivakoff}, {Stacey}, {Stutz}, {Stutzki},
  {Tahani}, {Thanjavur}, {Timmermann}, {Ullom}, {van Engelen}, {Vavagiakis},
  {Vissers}, {Wheeler}, {White}, {Zhu}, \& {Zou}}]{CPC2023}
{CCAT-Prime Collaboration}, {Aravena}, M., {Austermann}, J.~E., {et~al.} 2023,
  \href{http://dx.doi.org/10.3847/1538-4365/ac9838}{\color{blue}\apjs},
  \href{https://ui.adsabs.harvard.edu/abs/2023ApJS..264....7C}{264, 7}

\bibitem[{{Chapman} {et~al.}(2022){Chapman}, {Huber}, {Sinclair}, {Wheeler},
  {Austermann}, {Beall}, {Burgoyne}, {Choi}, {Crites}, {Duell}, {Devina},
  {Gao}, {Fich}, {Henke}, {Herter}, {Johnstone}, {Knee}, {Niemack}, {Rossi},
  {Stacey}, {Tsuchitori}, {Ullom}, {Van Lanen}, {Vavagiakis}, \&
  {Vissers}}]{Chapman2022}
{Chapman}, S.~C., {Huber}, A.~I., {Sinclair}, A.~K., {et~al.} 2022, in Society
  of Photo-Optical Instrumentation Engineers (SPIE) Conference Series, Vol.
  12190, Millimeter, Submillimeter, and Far-Infrared Detectors and
  Instrumentation for Astronomy XI, ed. J.~{Zmuidzinas} \& J.-R. {Gao},
  \href{https://ui.adsabs.harvard.edu/abs/2022SPIE12190E..05C}{1219005}

\bibitem[{{Chen} {et~al.}(2022){Chen}, {Zhang}, {Yang}, \& {Zheng}}]{Chen2022}
{Chen}, Z., {Zhang}, P., {Yang}, X., \& {Zheng}, Y. 2022,
  \href{http://dx.doi.org/10.1093/mnras/stab3604}{\color{blue}\mnras},
  \href{https://ui.adsabs.harvard.edu/abs/2022MNRAS.510.5916C}{510, 5916}

\bibitem[{{Chluba} {et~al.}(2012){Chluba}, {Nagai}, {Sazonov}, \&
  {Nelson}}]{Chluba2012}
{Chluba}, J., {Nagai}, D., {Sazonov}, S., \& {Nelson}, K. 2012,
  \href{http://dx.doi.org/10.1111/j.1365-2966.2012.21741.x}{\color{blue}\mnras},
  \href{https://ui.adsabs.harvard.edu/abs/2012MNRAS.426..510C}{426, 510}

\bibitem[{{Chluba} {et~al.}(2013){Chluba}, {Switzer}, {Nelson}, \&
  {Nagai}}]{Chluba2013}
{Chluba}, J., {Switzer}, E., {Nelson}, K., \& {Nagai}, D. 2013,
  \href{http://dx.doi.org/10.1093/mnras/stt110}{\color{blue}\mnras},
  \href{https://ui.adsabs.harvard.edu/abs/2013MNRAS.430.3054C}{430, 3054}

\bibitem[{{Comerford} \& {Natarajan}(2007)}]{Comerford2007}
{Comerford}, J.~M. \& {Natarajan}, P. 2007,
  \href{http://dx.doi.org/10.1111/j.1365-2966.2007.11934.x}{\color{blue}\mnras},
  \href{https://ui.adsabs.harvard.edu/abs/2007MNRAS.379..190C}{379, 190}

\bibitem[{{Cort{\'e}s} {et~al.}(2020){Cort{\'e}s}, {Cort{\'e}s}, {Reeves},
  {Bustos}, \& {Radford}}]{Cortes2020}
{Cort{\'e}s}, F., {Cort{\'e}s}, K., {Reeves}, R., {Bustos}, R., \& {Radford},
  S. 2020,
  \href{http://dx.doi.org/10.1051/0004-6361/202037784}{\color{blue}\aap},
  \href{https://ui.adsabs.harvard.edu/abs/2020A&A...640A.126C}{640, A126}

\bibitem[{{Dicker} {et~al.}(2014){Dicker}, {Ade}, {Aguirre}, {Brevik}, {Cho},
  {Datta}, {Devlin}, {Dober}, {Egan}, {Ford}, {Ford}, {Hilton}, {Hubmayr},
  {Irwin}, {Mason}, {Marganian}, {Mello}, {McMahon}, {Mroczkowski}, {Romero},
  {Stanchfield}, {Tucker}, {Vale}, {White}, {Whitehead}, \&
  {Young}}]{Dicker2014}
{Dicker}, S.~R., {Ade}, P.~A.~R., {Aguirre}, J., {et~al.} 2014, in Society of
  Photo-Optical Instrumentation Engineers (SPIE) Conference Series, Vol. 9153,
  Millimeter, Submillimeter, and Far-Infrared Detectors and Instrumentation for
  Astronomy VII, ed. W.~S. {Holland} \& J.~{Zmuidzinas},
  \href{https://ui.adsabs.harvard.edu/abs/2014SPIE.9153E..0JD}{91530J}

\bibitem[{{Du} {et~al.}(2015){Du}, {Fan}, {Shan}, {Zhao}, {Covone}, {Fu}, \&
  {Kneib}}]{Du2015}
{Du}, W., {Fan}, Z., {Shan}, H., {et~al.} 2015,
  \href{http://dx.doi.org/10.1088/0004-637X/814/2/120}{\color{blue}\apj},
  \href{https://ui.adsabs.harvard.edu/abs/2015ApJ...814..120D}{814, 120}

\bibitem[{{Eke} {et~al.}(2001){Eke}, {Navarro}, \& {Steinmetz}}]{Eke2001}
{Eke}, V.~R., {Navarro}, J.~F., \& {Steinmetz}, M. 2001,
  \href{http://dx.doi.org/10.1086/321345}{\color{blue}\apj},
  \href{https://ui.adsabs.harvard.edu/abs/2001ApJ...554..114E}{554, 114}

\bibitem[{{G{\'o}rski} {et~al.}(2005){G{\'o}rski}, {Hivon}, {Banday},
  {Wandelt}, {Hansen}, {Reinecke}, \& {Bartelmann}}]{Gorski2005}
{G{\'o}rski}, K.~M., {Hivon}, E., {Banday}, A.~J., {et~al.} 2005,
  \href{http://dx.doi.org/10.1086/427976}{\color{blue}\apj},
  \href{https://ui.adsabs.harvard.edu/abs/2005ApJ...622..759G}{622, 759}

\bibitem[{{Hand} {et~al.}(2012){Hand}, {Addison}, {Aubourg}, {Battaglia},
  {Battistelli}, {Bizyaev}, {Bond}, {Brewington}, {Brinkmann}, {Brown}, {Das},
  {Dawson}, {Devlin}, {Dunkley}, {Dunner}, {Eisenstein}, {Fowler}, {Gralla},
  {Hajian}, {Halpern}, {Hilton}, {Hincks}, {Hlozek}, {Hughes}, {Infante},
  {Irwin}, {Kosowsky}, {Lin}, {Malanushenko}, {Malanushenko}, {Marriage},
  {Marsden}, {Menanteau}, {Moodley}, {Niemack}, {Nolta}, {Oravetz}, {Page},
  {Palanque-Delabrouille}, {Pan}, {Reese}, {Schlegel}, {Schneider}, {Sehgal},
  {Shelden}, {Sievers}, {Sif{\'o}n}, {Simmons}, {Snedden}, {Spergel}, {Staggs},
  {Swetz}, {Switzer}, {Trac}, {Weaver}, {Wollack}, {Yeche}, \&
  {Zunckel}}]{Hand2012}
{Hand}, N., {Addison}, G.~E., {Aubourg}, E., {et~al.} 2012,
  \href{http://dx.doi.org/10.1103/PhysRevLett.109.041101}{\color{blue}\prl},
  \href{https://ui.adsabs.harvard.edu/abs/2012PhRvL.109d1101H}{109, 041101}

\bibitem[{{Hauser} {et~al.}(1998){Hauser}, {Arendt}, {Kelsall}, {Dwek},
  {Odegard}, {Weiland}, {Freudenreich}, {Reach}, {Silverberg}, {Moseley},
  {Pei}, {Lubin}, {Mather}, {Shafer}, {Smoot}, {Weiss}, {Wilkinson}, \&
  {Wright}}]{Hauser1998}
{Hauser}, M.~G., {Arendt}, R.~G., {Kelsall}, T., {et~al.} 1998,
  \href{http://dx.doi.org/10.1086/306379}{\color{blue}\apj},
  \href{https://ui.adsabs.harvard.edu/abs/1998ApJ...508...25H}{508, 25}

\bibitem[{{Hauser} \& {Dwek}(2001)}]{Hauser2001}
{Hauser}, M.~G. \& {Dwek}, E. 2001,
  \href{http://dx.doi.org/10.1146/annurev.astro.39.1.249}{\color{blue}\araa},
  \href{https://ui.adsabs.harvard.edu/abs/2001ARA&A..39..249H}{39, 249}

\bibitem[{{He} {et~al.}(2022){He}, {Dent}, \& {Wilson}}]{He2022}
{He}, H., {Dent}, W. R.~F., \& {Wilson}, C. 2022,
  \href{http://dx.doi.org/10.1088/1538-3873/aca717}{\color{blue}\pasp},
  \href{https://ui.adsabs.harvard.edu/abs/2022PASP..134l5001H}{134, 125001}

\bibitem[{{Kravtsov} \& {Borgani}(2012)}]{Kravtsov2012}
{Kravtsov}, A.~V. \& {Borgani}, S. 2012,
  \href{http://dx.doi.org/10.1146/annurev-astro-081811-125502}{\color{blue}\araa},
  \href{https://ui.adsabs.harvard.edu/abs/2012ARA&A..50..353K}{50, 353}

\bibitem[{{Lai} {et~al.}(2023){Lai}, {Howlett}, \& {Davis}}]{Lai2023}
{Lai}, Y., {Howlett}, C., \& {Davis}, T.~M. 2023,
  \href{http://dx.doi.org/10.1093/mnras/stac3252}{\color{blue}\mnras},
  \href{https://ui.adsabs.harvard.edu/abs/2023MNRAS.518.1840L}{518, 1840}

\bibitem[{{Lee} {et~al.}(2018){Lee}, {Le Brun}, {Haq}, {Deering}, {King},
  {Applegate}, \& {McCarthy}}]{Lee2018}
{Lee}, B.~E., {Le Brun}, A.~M.~C., {Haq}, M.~E., {et~al.} 2018,
  \href{http://dx.doi.org/10.1093/mnras/sty1377}{\color{blue}\mnras},
  \href{https://ui.adsabs.harvard.edu/abs/2018MNRAS.479..890L}{479, 890}

\bibitem[{{Ludlow} {et~al.}(2012){Ludlow}, {Navarro}, {Li}, {Angulo},
  {Boylan-Kolchin}, \& {Bett}}]{Ludlow2012}
{Ludlow}, A.~D., {Navarro}, J.~F., {Li}, M., {et~al.} 2012,
  \href{http://dx.doi.org/10.1111/j.1365-2966.2012.21892.x}{\color{blue}\mnras},
  \href{https://ui.adsabs.harvard.edu/abs/2012MNRAS.427.1322L}{427, 1322}

\bibitem[{{Macci{\`o}} {et~al.}(2007){Macci{\`o}}, {Dutton}, {van den Bosch},
  {Moore}, {Potter}, \& {Stadel}}]{Maccio2007}
{Macci{\`o}}, A.~V., {Dutton}, A.~A., {van den Bosch}, F.~C., {et~al.} 2007,
  \href{http://dx.doi.org/10.1111/j.1365-2966.2007.11720.x}{\color{blue}\mnras},
  \href{https://ui.adsabs.harvard.edu/abs/2007MNRAS.378...55M}{378, 55}

\bibitem[{{Nagai} {et~al.}(2003){Nagai}, {Kravtsov}, \& {Kosowsky}}]{Nagai2003}
{Nagai}, D., {Kravtsov}, A.~V., \& {Kosowsky}, A. 2003,
  \href{http://dx.doi.org/10.1086/368281}{\color{blue}\apj},
  \href{https://ui.adsabs.harvard.edu/abs/2003ApJ...587..524N}{587, 524}

\bibitem[{{Nagai} {et~al.}(2007){Nagai}, {Vikhlinin}, \&
  {Kravtsov}}]{Nagai2007}
{Nagai}, D., {Vikhlinin}, A., \& {Kravtsov}, A.~V. 2007,
  \href{http://dx.doi.org/10.1086/509868}{\color{blue}\apj},
  \href{https://ui.adsabs.harvard.edu/abs/2007ApJ...655...98N}{655, 98}

\bibitem[{{Navarro} {et~al.}(1996){Navarro}, {Frenk}, \& {White}}]{Navarro1996}
{Navarro}, J.~F., {Frenk}, C.~S., \& {White}, S. D.~M. 1996,
  \href{http://dx.doi.org/10.1086/177173}{\color{blue}\apj},
  \href{https://ui.adsabs.harvard.edu/abs/1996ApJ...462..563N}{462, 563}

\bibitem[{{Navarro} {et~al.}(1997){Navarro}, {Frenk}, \& {White}}]{Navarro1997}
{Navarro}, J.~F., {Frenk}, C.~S., \& {White}, S. D.~M. 1997,
  \href{http://dx.doi.org/10.1086/304888}{\color{blue}\apj},
  \href{https://ui.adsabs.harvard.edu/abs/1997ApJ...490..493N}{490, 493}

\bibitem[{{Nelson} {et~al.}(2019){Nelson}, {Springel}, {Pillepich},
  {Rodriguez-Gomez}, {Torrey}, {Genel}, {Vogelsberger}, {Pakmor}, {Marinacci},
  {Weinberger}, {Kelley}, {Lovell}, {Diemer}, \& {Hernquist}}]{Nelson2019}
{Nelson}, D., {Springel}, V., {Pillepich}, A., {et~al.} 2019,
  \href{http://dx.doi.org/10.1186/s40668-019-0028-x}{\color{blue}Computational
  Astrophysics and Cosmology},
  \href{https://ui.adsabs.harvard.edu/abs/2019ComAC...6....2N}{6, 2}

\bibitem[{{Nozawa} {et~al.}(2006){Nozawa}, {Itoh}, {Suda}, \&
  {Ohhata}}]{Nozawa2006}
{Nozawa}, S., {Itoh}, N., {Suda}, Y., \& {Ohhata}, Y. 2006,
  \href{http://dx.doi.org/10.1393/ncb/i2005-10223-0}{\color{blue}Nuovo Cimento
  B Serie}, \href{https://ui.adsabs.harvard.edu/abs/2006NCimB.121..487N}{121,
  487}

\bibitem[{{Okumura} \& {Taruya}(2022)}]{Okumura2022}
{Okumura}, T. \& {Taruya}, A. 2022,
  \href{http://dx.doi.org/10.1103/PhysRevD.106.043523}{\color{blue}\prd},
  \href{https://ui.adsabs.harvard.edu/abs/2022PhRvD.106d3523O}{106, 043523}

\bibitem[{{Pakmor} {et~al.}(2016){Pakmor}, {Springel}, {Bauer}, {Mocz},
  {Munoz}, {Ohlmann}, {Schaal}, \& {Zhu}}]{Pakmor2016}
{Pakmor}, R., {Springel}, V., {Bauer}, A., {et~al.} 2016,
  \href{http://dx.doi.org/10.1093/mnras/stv2380}{\color{blue}\mnras},
  \href{https://ui.adsabs.harvard.edu/abs/2016MNRAS.455.1134P}{455, 1134}

\bibitem[{{Parshley} {et~al.}(2018){Parshley}, {Kronshage}, {Blair}, {Herter},
  {Nolta}, {Stacey}, {Bazarko}, {Bertoldi}, {Bustos}, {Campbell}, {Chapman},
  {Cothard}, {Devlin}, {Erler}, {Fich}, {Gallardo}, {Giovanelli}, {Graf},
  {Gramke}, {Haynes}, {Hills}, {Limon}, {Mangum}, {McMahon}, {Niemack},
  {Nikola}, {Omlor}, {Riechers}, {Steeger}, {Stutzki}, \&
  {Vavagiakis}}]{Parshley2018}
{Parshley}, S.~C., {Kronshage}, J., {Blair}, J., {et~al.} 2018, in Society of
  Photo-Optical Instrumentation Engineers (SPIE) Conference Series, Vol. 10700,
  Ground-based and Airborne Telescopes VII, ed. H.~K. {Marshall} \&
  J.~{Spyromilio},
  \href{https://ui.adsabs.harvard.edu/abs/2018SPIE10700E..5XP}{107005X}

\bibitem[{{Planck Collaboration} {et~al.}(2014){Planck Collaboration}, {Ade},
  {Aghanim}, {Armitage-Caplan}, {Arnaud}, {Ashdown}, {Atrio-Barandela},
  {Aumont}, {Baccigalupi}, {Banday}, {Barreiro}, {Bartlett}, {Battaner},
  {Benabed}, {Beno{\^\i}t}, {Benoit-L{\'e}vy}, {Bernard}, {Bersanelli},
  {Bethermin}, {Bielewicz}, {Blagrave}, {Bobin}, {Bock}, {Bonaldi}, {Bond},
  {Borrill}, {Bouchet}, {Boulanger}, {Bridges}, {Bucher}, {Burigana}, {Butler},
  {Cardoso}, {Catalano}, {Challinor}, {Chamballu}, {Chen}, {Chiang}, {Chiang},
  {Christensen}, {Church}, {Clements}, {Colombi}, {Colombo}, {Couchot},
  {Coulais}, {Crill}, {Curto}, {Cuttaia}, {Danese}, {Davies}, {Davis}, {de
  Bernardis}, {de Rosa}, {de Zotti}, {Delabrouille}, {Delouis}, {D{\'e}sert},
  {Dickinson}, {Diego}, {Dole}, {Donzelli}, {Dor{\'e}}, {Douspis}, {Dupac},
  {Efstathiou}, {En{\ss}lin}, {Eriksen}, {Finelli}, {Forni}, {Frailis},
  {Franceschi}, {Galeotta}, {Ganga}, {Ghosh}, {Giard}, {Giraud-H{\'e}raud},
  {Gonz{\'a}lez-Nuevo}, {G{\'o}rski}, {Gratton}, {Gregorio}, {Gruppuso},
  {Hansen}, {Hanson}, {Harrison}, {Helou}, {Henrot-Versill{\'e}},
  {Hern{\'a}ndez-Monteagudo}, {Herranz}, {Hildebrandt}, {Hivon}, {Hobson},
  {Holmes}, {Hornstrup}, {Hovest}, {Huffenberger}, {Jaffe}, {Jaffe}, {Jones},
  {Juvela}, {Kalberla}, {Keih{\"a}nen}, {Kerp}, {Keskitalo}, {Kisner},
  {Kneissl}, {Knoche}, {Knox}, {Kunz}, {Kurki-Suonio}, {Lacasa}, {Lagache},
  {L{\"a}hteenm{\"a}ki}, {Lamarre}, {Langer}, {Lasenby}, {Laureijs},
  {Lawrence}, {Leonardi}, {Le{\'o}n-Tavares}, {Lesgourgues}, {Liguori},
  {Lilje}, {Linden-V{\o}rnle}, {L{\'o}pez-Caniego}, {Lubin},
  {Mac{\'\i}as-P{\'e}rez}, {Maffei}, {Maino}, {Mandolesi}, {Maris}, {Marshall},
  {Martin}, {Mart{\'\i}nez-Gonz{\'a}lez}, {Masi}, {Massardi}, {Matarrese},
  {Matthai}, {Mazzotta}, {Melchiorri}, {Mendes}, {Mennella}, {Migliaccio},
  {Mitra}, {Miville-Desch{\^e}nes}, {Moneti}, {Montier}, {Morgante},
  {Mortlock}, {Munshi}, {Murphy}, {Naselsky}, {Nati}, {Natoli}, {Netterfield},
  {N{\o}rgaard-Nielsen}, {Noviello}, {Novikov}, {Novikov}, {Osborne},
  {Oxborrow}, {Paci}, {Pagano}, {Pajot}, {Paladini}, {Paoletti}, {Partridge},
  {Pasian}, {Patanchon}, {Perdereau}, {Perotto}, {Perrotta}, {Piacentini},
  {Piat}, {Pierpaoli}, {Pietrobon}, {Plaszczynski}, {Pointecouteau}, {Polenta},
  {Ponthieu}, {Popa}, {Poutanen}, {Pratt}, {Pr{\'e}zeau}, {Prunet}, {Puget},
  {Rachen}, {Reach}, {Rebolo}, {Reinecke}, {Remazeilles}, {Renault},
  {Ricciardi}, {Riller}, {Ristorcelli}, {Rocha}, {Rosset}, {Roudier},
  {Rowan-Robinson}, \& {Rubi{\~n}o-Mart{\'\i}n}}]{PlanckCollaboration2014}
{Planck Collaboration}, {Ade}, P.~A.~R., {Aghanim}, N., {et~al.} 2014,
  \href{http://dx.doi.org/10.1051/0004-6361/201322093}{\color{blue}\aap},
  \href{https://ui.adsabs.harvard.edu/abs/2014A&A...571A..30P}{571, A30}

\bibitem[{{Planck Collaboration} {et~al.}(2016{\natexlab{a}}){Planck
  Collaboration}, {Ade}, {Aghanim}, {Arnaud}, {Ashdown}, {Aubourg}, {Aumont},
  {Baccigalupi}, {Banday}, {Barreiro}, {Bartolo}, {Battaner}, {Benabed},
  {Benoit-L{\'e}vy}, {Bersanelli}, {Bielewicz}, {Bock}, {Bonaldi}, {Bonavera},
  {Bond}, {Borrill}, {Bouchet}, {Burigana}, {Calabrese}, {Cardoso}, {Catalano},
  {Chamballu}, {Chiang}, {Christensen}, {Clements}, {Colombo}, {Combet},
  {Crill}, {Curto}, {Cuttaia}, {Danese}, {Davies}, {Davis}, {de Bernardis}, {de
  Zotti}, {Delabrouille}, {Dickinson}, {Diego}, {Dolag}, {Donzelli},
  {Dor{\'e}}, {Douspis}, {Ducout}, {Dupac}, {Efstathiou}, {Elsner},
  {En{\ss}lin}, {Eriksen}, {Finelli}, {Forni}, {Frailis}, {Fraisse},
  {Franceschi}, {Frejsel}, {Galeotta}, {Galli}, {Ganga}, {G{\'e}nova-Santos},
  {Giard}, {Gjerl{\o}w}, {Gonz{\'a}lez-Nuevo}, {G{\'o}rski}, {Gregorio},
  {Gruppuso}, {Hansen}, {Harrison}, {Henrot-Versill{\'e}},
  {Hern{\'a}ndez-Monteagudo}, {Herranz}, {Hildebrandt}, {Hivon}, {Hobson},
  {Hornstrup}, {Huffenberger}, {Hurier}, {Jaffe}, {Jaffe}, {Jones}, {Juvela},
  {Keih{\"a}nen}, {Keskitalo}, {Kitaura}, {Kneissl}, {Knoche}, {Kunz},
  {Kurki-Suonio}, {Lagache}, {Lamarre}, {Lasenby}, {Lattanzi}, {Lawrence},
  {Leonardi}, {Le{\'o}n-Tavares}, {Levrier}, {Liguori}, {Lilje},
  {Linden-V{\o}rnle}, {L{\'o}pez-Caniego}, {Lubin}, {Ma},
  {Mac{\'\i}as-P{\'e}rez}, {Maffei}, {Maino}, {Mak}, {Mandolesi}, {Mangilli},
  {Maris}, {Martin}, {Mart{\'\i}nez-Gonz{\'a}lez}, {Masi}, {Matarrese},
  {McGehee}, {Melchiorri}, {Mennella}, {Migliaccio}, {Miville-Desch{\^e}nes},
  {Moneti}, {Montier}, {Morgante}, {Mortlock}, {Munshi}, {Murphy}, {Naselsky},
  {Nati}, {Natoli}, {Noviello}, {Novikov}, {Novikov}, {Oxborrow}, {Pagano},
  {Pajot}, {Paoletti}, {Perdereau}, {Perotto}, {Pettorino}, {Piacentini},
  {Piat}, {Pierpaoli}, {Pointecouteau}, {Polenta}, {Ponthieu}, {Pratt},
  {Puget}, {Puisieux}, {Rachen}, {Racine}, {Reach}, {Reinecke}, {Remazeilles},
  {Renault}, {Renzi}, {Ristorcelli}, {Rocha}, {Rosset}, {Rossetti}, {Roudier},
  {Rubi{\~n}o-Mart{\'\i}n}, {Rusholme}, {Sandri}, {Santos}, {Savelainen},
  {Savini}, {Scott}, {Spencer}, {Stolyarov}, {Sudiwala}, {Sunyaev}, {Sutton},
  {Suur-Uski}, {Sygnet}, {Tauber}, {Terenzi}, {Toffolatti}, {Tomasi}, {Tucci},
  {Valenziano}, {Valiviita}, {Van Tent}, {Vielva}, {Villa}, {Wade}, {Wandelt},
  {Wang}, {Wehus}, {Yvon}, {Zacchei}, \& {Zonca}}]{PlanckCollaboration2016a}
{Planck Collaboration}, {Ade}, P.~A.~R., {Aghanim}, N., {et~al.}
  2016{\natexlab{a}},
  \href{http://dx.doi.org/10.1051/0004-6361/201526328}{\color{blue}\aap},
  \href{https://ui.adsabs.harvard.edu/abs/2016A&A...586A.140P}{586, A140}

\bibitem[{{Planck Collaboration} {et~al.}(2016{\natexlab{b}}){Planck
  Collaboration}, {Ade}, {Aghanim}, {Arnaud}, {Ashdown}, {Aumont},
  {Baccigalupi}, {Banday}, {Barreiro}, {Bartlett}, {Bartolo}, {Battaner},
  {Battye}, {Benabed}, {Beno{\^\i}t}, {Benoit-L{\'e}vy}, {Bernard},
  {Bersanelli}, {Bielewicz}, {Bock}, {Bonaldi}, {Bonavera}, {Bond}, {Borrill},
  {Bouchet}, {Bucher}, {Burigana}, {Butler}, {Calabrese}, {Cardoso},
  {Catalano}, {Challinor}, {Chamballu}, {Chary}, {Chiang}, {Christensen},
  {Church}, {Clements}, {Colombi}, {Colombo}, {Combet}, {Comis}, {Couchot},
  {Coulais}, {Crill}, {Curto}, {Cuttaia}, {Danese}, {Davies}, {Davis}, {de
  Bernardis}, {de Rosa}, {de Zotti}, {Delabrouille}, {D{\'e}sert}, {Diego},
  {Dolag}, {Dole}, {Donzelli}, {Dor{\'e}}, {Douspis}, {Ducout}, {Dupac},
  {Efstathiou}, {Elsner}, {En{\ss}lin}, {Eriksen}, {Falgarone}, {Fergusson},
  {Finelli}, {Forni}, {Frailis}, {Fraisse}, {Franceschi}, {Frejsel},
  {Galeotta}, {Galli}, {Ganga}, {Giard}, {Giraud-H{\'e}raud}, {Gjerl{\o}w},
  {Gonz{\'a}lez-Nuevo}, {G{\'o}rski}, {Gratton}, {Gregorio}, {Gruppuso},
  {Gudmundsson}, {Hansen}, {Hanson}, {Harrison}, {Henrot-Versill{\'e}},
  {Hern{\'a}ndez-Monteagudo}, {Herranz}, {Hildebrandt}, {Hivon}, {Hobson},
  {Holmes}, {Hornstrup}, {Hovest}, {Huffenberger}, {Hurier}, {Jaffe}, {Jaffe},
  {Jones}, {Juvela}, {Keih{\"a}nen}, {Keskitalo}, {Kisner}, {Kneissl},
  {Knoche}, {Kunz}, {Kurki-Suonio}, {Lagache}, {L{\"a}hteenm{\"a}ki},
  {Lamarre}, {Lasenby}, {Lattanzi}, {Lawrence}, {Leonardi}, {Lesgourgues},
  {Levrier}, {Liguori}, {Lilje}, {Linden-V{\o}rnle}, {L{\'o}pez-Caniego},
  {Lubin}, {Mac{\'\i}as-P{\'e}rez}, {Maggio}, {Maino}, {Mandolesi}, {Mangilli},
  {Maris}, {Martin}, {Mart{\'\i}nez-Gonz{\'a}lez}, {Masi}, {Matarrese},
  {McGehee}, {Meinhold}, {Melchiorri}, {Melin}, {Mendes}, {Mennella},
  {Migliaccio}, {Mitra}, {Miville-Desch{\^e}nes}, {Moneti}, {Montier},
  {Morgante}, {Mortlock}, {Moss}, {Munshi}, {Murphy}, {Naselsky}, {Nati},
  {Natoli}, {Netterfield}, {N{\o}rgaard-Nielsen}, {Noviello}, {Novikov},
  {Novikov}, {Oxborrow}, {Paci}, {Pagano}, {Pajot}, {Paoletti}, {Partridge},
  {Pasian}, {Patanchon}, {Pearson}, {Perdereau}, {Perotto}, {Perrotta},
  {Pettorino}, {Piacentini}, {Piat}, {Pierpaoli}, {Pietrobon}, {Plaszczynski},
  {Pointecouteau}, {Polenta}, {Popa}, {Pratt}, {Pr{\'e}zeau}, {Prunet},
  {Puget}, {Rachen}, {Rebolo}, {Reinecke}, {Remazeilles}, {Renault}, {Renzi},
  {Ristorcelli}, {Rocha}, {Roman}, {Rosset}, {Rossetti}, {Roudier},
  {Rubi{\~n}o-Mart{\'\i}n}, {Rusholme}, {Sandri}, {Santos}, {Savelainen},
  {Savini}, {Scott}, {Seiffert}, {Shellard}, {Spencer}, {Stolyarov}, {Stompor},
  {Sudiwala}, {Sunyaev}, {Sutton}, {Suur-Uski}, {Sygnet}, {Tauber}, {Terenzi},
  {Toffolatti}, {Tomasi}, {Tristram}, {Tucci}, {Tuovinen}, {T{\"u}rler},
  {Umana}, {Valenziano}, {Valiviita}, {Van Tent}, {Vielva}, {Villa}, {Wade},
  {Wandelt}, {Wehus}, {Weller}, {White}, {Yvon}, {Zacchei}, \&
  {Zonca}}]{PlanckCollaboration2016}
{Planck Collaboration}, {Ade}, P.~A.~R., {Aghanim}, N., {et~al.}
  2016{\natexlab{b}},
  \href{http://dx.doi.org/10.1051/0004-6361/201525833}{\color{blue}\aap},
  \href{https://ui.adsabs.harvard.edu/abs/2016A&A...594A..24P}{594, A24}

\bibitem[{{Ricker} \& {Sarazin}(2001)}]{Ricker2001}
{Ricker}, P.~M. \& {Sarazin}, C.~L. 2001,
  \href{http://dx.doi.org/10.1086/323365}{\color{blue}\apj},
  \href{https://ui.adsabs.harvard.edu/abs/2001ApJ...561..621R}{561, 621}

\bibitem[{{Sayers} {et~al.}(2013){Sayers}, {Czakon}, {Mantz}, {Golwala},
  {Ameglio}, {Downes}, {Koch}, {Lin}, {Maughan}, {Molnar}, {Moustakas},
  {Mroczkowski}, {Pierpaoli}, {Shitanishi}, {Siegel}, {Umetsu}, \& {Van der
  Pyl}}]{Sayers2013}
{Sayers}, J., {Czakon}, N.~G., {Mantz}, A., {et~al.} 2013,
  \href{http://dx.doi.org/10.1088/0004-637X/768/2/177}{\color{blue}\apj},
  \href{https://ui.adsabs.harvard.edu/abs/2013ApJ...768..177S}{768, 177}

\bibitem[{{Sereno} \& {Covone}(2013)}]{Sereno2013}
{Sereno}, M. \& {Covone}, G. 2013,
  \href{http://dx.doi.org/10.1093/mnras/stt1086}{\color{blue}\mnras},
  \href{https://ui.adsabs.harvard.edu/abs/2013MNRAS.434..878S}{434, 878}

\bibitem[{{Shi} {et~al.}(2024){Shi}, {Zhang}, {Mao}, \& {Gu}}]{Shi2024}
{Shi}, Y., {Zhang}, P., {Mao}, S., \& {Gu}, Q. 2024,
  \href{http://dx.doi.org/10.1093/mnras/stae274}{\color{blue}\mnras}
  \href{https://ui.adsabs.harvard.edu/abs/2024MNRAS.tmp..296S}{[\eprint[arXiv]{2401.13871}]}

\bibitem[{{Springel}(2010)}]{Springel2010}
{Springel}, V. 2010,
  \href{http://dx.doi.org/10.1111/j.1365-2966.2009.15715.x}{\color{blue}\mnras},
  \href{https://ui.adsabs.harvard.edu/abs/2010MNRAS.401..791S}{401, 791}

\bibitem[{{Sun} \& {Murray}(2002)}]{Sun2002}
{Sun}, M. \& {Murray}, S.~S. 2002,
  \href{http://dx.doi.org/10.1086/341756}{\color{blue}\apj},
  \href{https://ui.adsabs.harvard.edu/abs/2002ApJ...576..708S}{576, 708}

\bibitem[{{Sunyaev} \& {Zeldovich}(1972)}]{Sunyaev1972}
{Sunyaev}, R.~A. \& {Zeldovich}, Y.~B. 1972, Comments on Astrophysics and Space
  Physics, \href{https://ui.adsabs.harvard.edu/abs/1972CoASP...4..173S}{4, 173}

\bibitem[{{Sunyaev} \& {Zeldovich}(1980)}]{Sunyaev1980}
{Sunyaev}, R.~A. \& {Zeldovich}, Y.~B. 1980,
  \href{http://dx.doi.org/10.1093/mnras/190.3.413}{\color{blue}\mnras},
  \href{https://ui.adsabs.harvard.edu/abs/1980MNRAS.190..413S}{190, 413}

\bibitem[{{Tanimura} {et~al.}(2021){Tanimura}, {Zaroubi}, \&
  {Aghanim}}]{Tanimura2021}
{Tanimura}, H., {Zaroubi}, S., \& {Aghanim}, N. 2021,
  \href{http://dx.doi.org/10.1051/0004-6361/202038846}{\color{blue}\aap},
  \href{https://ui.adsabs.harvard.edu/abs/2021A&A...645A.112T}{645, A112}

\bibitem[{{Terry} {et~al.}(2019){Terry}, {Battaglia}, {Basu}, {Beringue},
  {Bertoldi}, {Chapman}, {Choi}, {Cothard}, {Chung}, {Erler}, {Fich},
  {Foreman}, {Gallardo}, {Gao}, {Graf}, {Haynes}, {Herter}, {Hilton},
  {Hubmayr}, {Johnstone}, {Komatsu}, {Magnelli}, {Mauskopf}, {McMahon},
  {Meerburg}, {Meyers}, {Mittal}, {Niemack}, {Nikola}, {Parshley}, {Riechers},
  {Stacey}, {Stutzki}, {Vavagiakis}, {Viero}, \& {Vissers}}]{Terry2019}
{Terry}, H., {Battaglia}, N., {Basu}, K., {et~al.} 2019, in Bulletin of the
  American Astronomical Society, Vol.~51,
  \href{https://ui.adsabs.harvard.edu/abs/2019BAAS...51g.213T}{213}

\bibitem[{{Tucci} {et~al.}(2011){Tucci}, {Toffolatti}, {de Zotti}, \&
  {Mart{\'\i}nez-Gonz{\'a}lez}}]{Tucci2011}
{Tucci}, M., {Toffolatti}, L., {de Zotti}, G., \& {Mart{\'\i}nez-Gonz{\'a}lez},
  E. 2011,
  \href{http://dx.doi.org/10.1051/0004-6361/201116972}{\color{blue}\aap},
  \href{https://ui.adsabs.harvard.edu/abs/2011A&A...533A..57T}{533, A57}

\bibitem[{{Turner} {et~al.}(2023){Turner}, {Blake}, \& {Ruggeri}}]{Turner2023}
{Turner}, R.~J., {Blake}, C., \& {Ruggeri}, R. 2023,
  \href{http://dx.doi.org/10.1093/mnras/stac3256}{\color{blue}\mnras},
  \href{https://ui.adsabs.harvard.edu/abs/2023MNRAS.518.2436T}{518, 2436}

\bibitem[{{van Marrewijk} {et~al.}(2024){van Marrewijk}, {Di Mascolo}, {Gill},
  {Battaglia}, {Battistelli}, {Bond}, {Devlin}, {Doze}, {Dunkley}, {Knowles},
  {Hincks}, {Hughes}, {Hilton}, {Moodley}, {Mroczkowski}, {Naess}, {Partridge},
  {Popping}, {Sif{\'o}n}, {Staggs}, \& {Wollack}}]{vanMarrewijk2024}
{van Marrewijk}, J., {Di Mascolo}, L., {Gill}, A.~S., {et~al.} 2024,
  \href{http://dx.doi.org/10.1051/0004-6361/202348213}{\color{blue}\aap},
  \href{https://ui.adsabs.harvard.edu/abs/2024A&A...689A..41V}{689, A41}

\bibitem[{{Wang} {et~al.}(2021){Wang}, {Ramachandra}, {Salazar-Canizales},
  {Feldman}, {Watkins}, \& {Dolag}}]{Wang2021}
{Wang}, Y., {Ramachandra}, N., {Salazar-Canizales}, E.~M., {et~al.} 2021,
  \href{http://dx.doi.org/10.1093/mnras/stab1715}{\color{blue}\mnras},
  \href{https://ui.adsabs.harvard.edu/abs/2021MNRAS.506.1427W}{506, 1427}

\bibitem[{{Yao} {et~al.}(2023){Yao}, {Chen}, \& {Wang}}]{Yao2023}
{Yao}, Y.-W., {Chen}, W.-R., \& {Wang}, Z. 2023,
  \href{http://dx.doi.org/10.1088/1674-4527/acc29d}{\color{blue}Research in
  Astronomy and Astrophysics},
  \href{https://ui.adsabs.harvard.edu/abs/2023RAA....23d5013Y}{23, 045013}

\bibitem[{{Yuan} \& {Han}(2020)}]{Yuan2020}
{Yuan}, Z.~S. \& {Han}, J.~L. 2020,
  \href{http://dx.doi.org/10.1093/mnras/staa2363}{\color{blue}\mnras},
  \href{https://ui.adsabs.harvard.edu/abs/2020MNRAS.497.5485Y}{497, 5485}

\bibitem[{{Zhao} {et~al.}(2003){Zhao}, {Jing}, {Mo}, \&
  {B{\"o}rner}}]{Zhao2003}
{Zhao}, D.~H., {Jing}, Y.~P., {Mo}, H.~J., \& {B{\"o}rner}, G. 2003,
  \href{http://dx.doi.org/10.1086/379734}{\color{blue}\apjl},
  \href{https://ui.adsabs.harvard.edu/abs/2003ApJ...597L...9Z}{597, L9}

\end{thebibliography}

\end{document}